\documentclass{aa}

\usepackage{graphicx}
\usepackage{txfonts}
\usepackage{bbm}
\usepackage{url}
\usepackage{natbib}
\usepackage[utf8]{inputenc}
\usepackage[breaklinks=true]{hyperref} 
\hypersetup{colorlinks=true,urlcolor=blue,citecolor=blue,pdfborder= 0 0 0}
\bibpunct{(}{)}{;}{a}{}{,}    
\begin{document}

   \title{Bayesian group finder based on marked point processes}

   \subtitle{Method and feasibility study using the 2MRS data set}

   \author{Elmo~Tempel\inst{1,2}
          \and
		  Maarja~Kruuse\inst{2}
		  \and
		  Rain~Kipper\inst{2}
		  \and
		  Taavi~Tuvikene\inst{2}
		  \and
		  Jenny~G.~Sorce\inst{1,3}
		  \and
          Radu~S.~Stoica\inst{4}
          }

   \institute{Leibniz-Institut f\"ur Astrophysik Potsdam (AIP), An der Sternwarte 16, 14482 Potsdam, Germany\\
              \email{elmo.tempel@ut.ee}
        \and
             Tartu Observatory, University of Tartu, Observatooriumi 1, 61602 T\~oravere, Estonia
		\and
			Univ Lyon, Univ Lyon1, Ens de Lyon, CNRS, Centre de Recherche Astrophysique de Lyon UMR5574, F-69230, Saint-Genis-Laval, France
		\and
			Universit\'e de Lorraine, Institut Elie Cartan de Lorraine, 54506 Vandoeuvre-l\'es-Nancy Cedex, France}

   \date{Received April 12, 2018; accepted July 10, 2018}
   \date{}

 
  \abstract
   {Galaxy groups and clusters are formidable cosmological probes. They permit the studying of the environmental effects on galaxy formation. A reliable detection of galaxy groups is an open problem and is important for ongoing and future cosmological surveys.}
   {We propose a probabilistic galaxy group detection algorithm based on marked point processes with interactions.}
   {The pattern of galaxy groups in a catalogue is seen as a random set of interacting objects. The positions and the interactions of these objects are governed by a probability density. The parameters of the probability density were chosen using a priori knowledge. The estimator of the unknown cluster pattern is given by the configuration of objects maximising the proposed probability density. Adopting the Bayesian framework, the proposed probability density is maximised using a simulated annealing (SA) algorithm. At fixed temperature, the SA algorithm is a Monte Carlo sampler of the probability density. Hence, the method provides ``for free'' additional information such as the probabilities that a point or two points in the observation domain belong to the cluster pattern, respectively. These supplementary tools allow the construction of tests and techniques to validate and to refine the detection result.}
   {To test the feasibility of the proposed methodology, we applied it to the well-studied 2MRS data set. Compared to previously published Friends-of-Friends (FoF) group finders, the proposed Bayesian group finder gives overall similar results. However for specific applications, like the reconstruction of the local Universe, the details of the grouping algorithms are important.}
   {The proposed Bayesian group finder is tested on a galaxy redshift survey, but more detailed analyses are needed to understand the actual capabilities of the algorithm regarding upcoming cosmological surveys. The presented mathematical framework permits adapting it easily for other data sets (in astronomy and in other fields of sciences). In cosmology, one promising application is the detection of galaxy groups in photometric galaxy redshift surveys, while taking into account the full photometric redshift posteriors.}


   \maketitle
%
\section{Introduction}

	Galaxy groups and clusters are one of the most widely used systems in cosmology. For example, they are used to assess the cluster mass function in observations \citep{2003ApJ...585..182B, 2003A&A...397...63H, 2003AJ....126.1677P, 2004ApJ...601..610V, 2007ApJ...657..183R, 2010MNRAS.407..533W, 2014A&A...566A...1T, 2016MNRAS.463.1666C}, to estimate the halo assembly bias while combining the weak lensing and galaxy groups \citep{2017MNRAS.468.3251D} or to measure the geometry of the Universe from weak lensing behind the galaxy groups \citep{2012ApJ...749..127T}. Galaxy groups are also used to study the environmental effects on galaxy evolution \citep{2009A&A...495...37T, 2011MNRAS.411..675S, 2012A&A...545A.104L, 2012MNRAS.422.1835S, 2013MNRAS.436...34C, 2013MNRAS.431..167R, 2014MNRAS.438..262P, 2015MNRAS.451.3249A, 2015ApJ...800...24K, 2016MNRAS.455.4013D, 2016A&A...590A..29P, 2017A&A...597A..86P}. Galaxy groups are key targets in the Galaxy And Mass Assembly (GAMA\footnote{\url{http://www.gama-survey.org}.}) survey to test the cosmological models and to connect galaxy evolution with environmental mechanisms \citep{2009A&G....50e..12D}. A successor of GAMA survey is the forthcoming Wide Area Vista Extragalactic Survey (WAVES\footnote{\url{https://wavesurvey.org}.}; \citealt{2016ASSP...42..205D}) that is part of the four-metre Multi-Object Spectroscopic Telescope (4MOST\footnote{\url{https://www.4most.eu}.}) consortium \citep{2016SPIE.9908E..1OD}. WAVES survey is designed to probe the evolution of galaxies and structure down to the smallest galaxy groups.
	
	Detection of galaxy groups and clusters in cosmological redshift surveys is a classic problem. The first galaxy group samples were constructed by \citet{1976ApJS...32..409T}, \citet{1982ApJ...257..423H}, and \citet{1983ApJS...52...61G}. In these papers, group construction is based on a single-linkage agglomerative clustering algorithm, which in astronomical community is labelled as Friends-of-Friends (or simply FoF) or percolation algorithm\footnote{We note that this procedure is widely used in astronomical community, mostly without an awareness of its widespread use in other fields of sciences.}. The first attempt to model the spatial clustering of galaxies was made by \citet{1952ApJ...116..144N}, where authors propose a mathematical model for galaxy clustering (see \citealt{Lawson:02} for an overview of spatial cluster modelling in mathematics).
	
	The FoF algorithm have been used to build galaxy group catalogues for most major astronomical surveys, including Sloan Digital Sky Survey \citep{2005ApJ...630..759M, 2006ApJS..167....1B, 2007A&A...474..783D, 2008A&A...479..927T, 2010A&A...514A.102T, 2012MNRAS.423.1583M, 2012A&A...540A.106T, 2014A&A...566A...1T, 2017A&A...602A.100T, 2016ApJS..225...23S}, GAMA \citep{2011MNRAS.416.2640R}, Millennium Galaxy and Group Catalogue \citep{2011MNRAS.416..727C}, and 2dF galaxy redshift survey \citep{2002MNRAS.335..216M, 2004MNRAS.348..866E, 2006AN....327..365T}. FoF algorithm has a long history in astronomy and it is actively used since \citet{1976ApJS...32..409T}. Some of the older surveys, where FoF algorithm was used are the Centre for Astrophysics redshift survey \citep{1987MNRAS.225..505N, 1989ApJ...344...57R, 1997AJ....113..483R}, Perseus-Pisces Survey \citep{1998A&AS..130..341T}, ESO slice project \citep{1999A&A...342....1R}, the Nearby Optical Galaxy sample \citep{2000ApJ...543..178G}, Las Campanas Redshift Survey \citep{1989ApJS...69..809M, 2000ApJS..130..237T}, Updated Zwicky Catalog \citep{2002AJ....123.2976R}, Southern Sky Redshift Survey \citep{2002A&A...381..420A}. Most of the previously mentioned surveys are local (redshift $z<0.2$), however, FoF algorithm has also been used to detect galaxy groups in intermediate redshift surveys \citep{2001ApJ...552..427C, 2005MNRAS.358...71W, 2009ApJ...697.1842K, 2012ApJ...753..121K} and in quasar catalogues \citep{2011MNRAS.417.1402F}.
	
	The key point in using the FoF algorithm is related to the choice of the linking length value\footnote{Linking length is a parameter in a FoF algorithm that is used to determine whether two galaxies belong to the same system. If distance between galaxies is smaller than a given linking length then these galaxies are linked together.} and how the linking length value depends on the distance and underlying galaxy density. Because of its simplicity the method is fast, easy to apply and simple to understand. However, it is not superior than other methods for cluster detection. The performance of the FoF method using different linking lengths is studied in \citet{2014MNRAS.440.1763D}, where they conclude that the optimal linking length values depend on the scientific goal of the group catalogue.
	
	Another group of methods for group and cluster detection are halo-based group finders \citep{2005MNRAS.356.1293Y, 2007ApJ...671..153Y, 2015MNRAS.453.3848D}. The idea of these methods is to use the underlying cosmological model and to group together galaxies that belong to the same dark matter halo. In addition to the FoF and halo-based group finders, there are other group finders that are occasionally used, including the Voronoi-Delaunay method \citep{2001A&A...368..776R, 2005ApJ...625....6G, 2010A&A...520A..42C, 2012ApJ...751...50G, 2017ApJ...838..109P}, matched filter techniques \citep{1999ApJ...517...78K, 2010MNRAS.406..673M, 2018MNRAS.473.5221B} and density field based methods \citep{2005AJ....130..968M, 2009ApJ...703.1061S, 2012MNRAS.422...25S}. \citet{2018ApJ...861...22A} propose a weighting technique to determine galaxy group and cluster membership. Another class of group finders are developed for photometric redshift surveys, which take advantage of the red sequence of galaxies \citep{2000AJ....120.2148G, 2007ApJ...660..221K, 2014ApJ...785..104R, 2016MNRAS.455.3020L}. An extensive comparison of various group detection methods is carried out by \citet{2014MNRAS.441.1513O, 2015MNRAS.449.1897O, 2018MNRAS.475..853O}. The reliability of different group detection methods is also assessed by \citet{1995ApJS...97..259F}, \citet{2013MNRAS.436..380N}, and \citet{2014MNRAS.440.1763D}.
   
	An interesting region for group detection is the local Universe that is covered by the Two Micron All Sky Survey \citep[2MASS;][]{2006AJ....131.1163S} Redshift Survey \citep[2MRS;][]{2012ApJS..199...26H}. The first catalogue of groups based on a FoF was constructed by \citet{2007ApJ...655..790C}. In recent years, several new group catalogues for 2MRS have been constructed \citep{2015AJ....149..171T, 2016ApJ...832...39L, 2016A&A...596A..14S, 2016A&A...588A..14T, 2017ApJ...843...16K, 2017MNRAS.470.2982L}. The 2MRS groups have been used in many studies, including the reconstruction of the local Universe via constrained simulations \citep{2017MNRAS.469.2859S, 2018MNRAS.476.4362S}, to measure the gas content of galaxy groups \citep{2013AJ....146..124H}, to analyse the galaxy properties in poor groups of galaxies \citep{2009ApJ...696.1441T} or to measure the intergalactic medium in fossil galaxy groups \citep{2014MNRAS.444..651M}.

	It is undeniable that galaxy groups are widely used entities in astronomy and the detection of galaxy groups from observed data sets is an important task. In coming decade several new observational surveys will be carried out that will allow us to map the galaxy group environment with an unprecedented detail. This will pose challenges to the available methods. The continuous increase of computational resources opens the possibility to develop new group finders that take advantage of the Bayesian approach.
 
	Most of the currently available methods for group and cluster detection in cosmology are driven by the desire to detect galaxy groups and clusters. The open question related to the previously referred methods is the following: does the detected clusters exist because of the method or because of the data? From mathematical point of view the detection of galaxy clusters is an open problem that is not solved yet.
 
	The current paper aims to integrate the cluster detection methodology based on marked point processes described in \citet{Stoica:04, 2005A&A...434..423S, Stoica:07, 2007JRSSC..56....1S, 2010A&A...510A..38S} within a new tool for group and cluster detection for cosmological redshift surveys. The method we propose for galaxy group detection is a Bayesian method developed for pattern detection that can be adapted to any data sets. Instead of focusing on the detection of points forming clusters, our proposed method detects the spatial regions where those points belong to. Our aim is to model the clustered pattern that is constrained by the observational data. The used Bayesian methodology allows morphological, quantitative and qualitative characterisation of the detected cluster pattern.
	
	In the current paper we introduce the new methodology in its ``simplest'' form and apply it to the 2MRS spectroscopic survey to demonstrate the feasibility of our developed Bayesian methodology. Full capabilities of our proposed method will be analysed in a separate study. A promising perspective is to apply the developed methodology to photometric redshift surveys utilising the full photometric redshift posteriors. This is especially promising for the currently ongoing Javalambre Physics of the Accelerating Universe Astrophysical Survey\footnote{\url{http://www.j-pas.org}.} \citep[J-PAS;][]{2014arXiv1403.5237B}. The proposed Bayesian group finder will potentially complement available group finders for photometric redshift surveys \citep{2008AJ....135..809L, 2008ApJ...681.1046L, 2011MNRAS.410...13G, 2014ApJ...788..109J, 2014A&A...561A..71Z, 2015MNRAS.452..549A}.
	
	The structure of the paper is following. In Section~\ref{sec:bayes_groups} we describe the Bayesian marked point process based methodology for galaxy group detection, in Section~\ref{sec:group_extraction} we give the algorithm to extract single galaxy groups using the detected probabilisitic clustered pattern. The proposed methodology is applied to the 2MRS data set in Section~\ref{sec:application} and the constructed group catalogue is presented in Section~\ref{sec:2mrs_groups}. In Section~\ref{sec:conclusions} we present our conclusions and discuss future prospects. Appendix~\ref{app:cat} gives the description of our 2MRS Bayesian group catalogue that is made available in \url{http://cosmodb.to.ee}. In Appendix~\ref{app:mock} we apply the proposed method to a simulated mock data set and analyse the completeness and contamination of detected groups and clusters. Throughout this paper we assume the Planck cosmology \citep{2016A&A...594A..13P}: the Hubble constant $H_0 = 67.8~\mathrm{km~s^{-1}Mpc^{-1}}$, the matter density $\Omega_\mathrm{m} = 0.308$, and the dark energy density $\Omega_\Lambda = 0.692$.

\section{Bayesian group finder based on marked point processes}
\label{sec:bayes_groups}

\subsection{Set-up of the problem}

The galaxy cluster pattern detection in cosmological observations has two challenges to be tackled. The first one is the definition of a cluster. It includes the following aspects. The clusters boundaries are not well defined and poor groups (with small number of galaxies) blend in with the underlying cosmic web, which complicates the detection of poor systems. Furthermore, the detection of galaxy groups and clusters in galaxy redshift surveys is complicated due to various observational selection effects. The second challenge to deal with is the definition of the cluster pattern. The latter together with the method to model the cluster pattern are described in Sections~\ref{sec:new_methodology}--\ref{sec:simulation}. In this section, we describe the observational data set and discuss some of the observational selection effects that significantly affect the group and cluster detection in cosmology.

Galaxy positions in galaxy redshift survey are spatial data, that is a point field $\mathcal{D}$ in an observational bounded region $W$. A galaxy position is described by three coordinates: two coordinates on the plane of the sky (spherical coordinates) and the third one gives the distance from the observer to the galaxy. Based on these three coordinates, cartesian coordinates can be computed. Since in observations, the distance to a galaxy is not a directly measurable parameter, the point field $\mathcal{D}$ is not isotropic and galaxies distributions along the line of sight and perpendicular to it are different. The galaxy distribution is elongated along the line of sight (see discussion below). The observational window $W$ defines the spatial region, where galaxies are observed. In sky plane observational window $W$ is defined by the footprint of a survey, in radial direction it is limited by minimum and maximum distances.

Galaxy cluster detection in redshift surveys is a cluster pattern detection problem in spatial data. Here a cluster is defined as a set of points that are grouped in a rather bounded region of the observation window, while the points characteristics exhibit ``similar statistical properties''. The cluster pattern is defined as the set of clusters. Cluster pattern detection means either finding the subset of points in $\mathcal{D}$ forming the cluster pattern, or detecting the geometrical characteristics of the regions underlined by the clusters in the observed data field (e.g. location, perimeter, volume). The  cluster pattern detection method we propose takes into account the following aspects: heterogeneity of the data, smoothing effects (e.g. observational uncertainties), statistical descriptors of the cluster pattern, probability of having a cluster in a given region, testing for cluster presence.

\begin{figure}
\centering
\includegraphics[width=88mm]{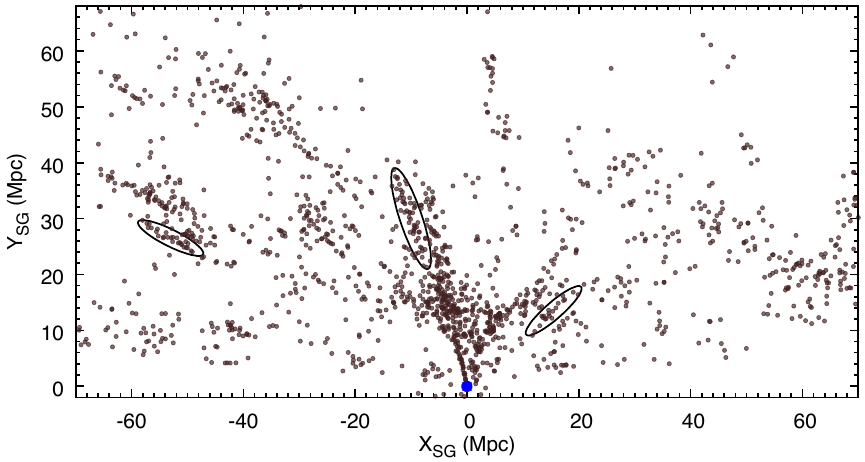}
	\caption{Distribution of galaxies (a point field $\mathcal{D}$) in the 2MRS data set. Positions of galaxies are given in supergalactic coordinates, where observer is located at the origin of coordinates (marked as blue point on the figure). The thickness of the slice shown in the figure is 15~Mpc. Some galaxy clusters are visually identified and marked with black ellipses to highlight the elongation of galaxy groups and clusters along the line of sight.}
	\label{fig:gal_field}
\end{figure}

The data point field $\mathcal{D}$ we analyse has the following specific properties. Only two galaxy coordinates (spherical sky coordinates) are known precisely, while third coordinate (distance) is only an approximation. Galaxy distances are measured using the observed recessional velocities (redshifts) and Hubble's law. Therefore, even if the redshift is precisely measured, the distance cannot be inferred accurately from it. The distance estimates of galaxies are affected by the peculiar motions of galaxies in the Universe. Due to the gravitational pull of galaxy groups and clusters, the peculiar velocities of galaxies are largest in galaxy clusters, hence the distance estimates of galaxies in galaxy clusters are most uncertain. As a result of this, all galaxy groups and clusters seem elongated along the line of sight in redshift-space as first noted by \citet{1972MNRAS.156P...1J}. Figure~\ref{fig:gal_field} illustrates the galaxy distribution in redshift surveys. The strong elongation of galaxy systems along the line of sight is clearly visible. This is important property of the observed galaxy field that we specifically include in our model (see Section~\ref{sec:new_methodology}).

In addition to the anisotropic distribution of galaxies, the galaxy distribution in cosmological redshift surveys is not homogeneous. Most redshift surveys are flux-limited surveys, meaning that only galaxies brighter than some limiting observed magnitude (luminosity) are observed. Since the observed luminosity of galaxies decreases as a function of distance, it poses a distance dependent intrinsic luminosity limit. Moreover, most of the galaxies in the Universe are faint galaxies \citep[see the luminosity function of galaxies, e.g.][]{2011A&A...529A..53T}, which means that the number density of galaxies can decrease tenfold in a redshift survey within a meaningful distance interval \citep[see e.g.][]{2012A&A...539A..80L}. In practice, it means that the decrease of number density of galaxies (points) as a function of distance should be taken into account by the cluster pattern detection algorithm.

\subsection{Principles of the cluster pattern detection method based on marked point processes}
\label{sec:new_methodology}

In most widely used algorithms (e.g. FoF), the group and cluster finding algorithm focuses on the detection of galaxies forming the clusters. The methodology presented in this paper adopts a different point of view. Our method detects the spatial regions where galaxies belong to. According to this, galaxy cluster is defined by the following properties:
\begin{itemize}
	\item the galaxies are grouped in a rather bounded region, while their characteristics share the ``similar statistical properties'',
	\item the region occupied by a cluster is approximated by the set-theoretic union of a finite number of overlapping compact objects with random centres and shape parameters.
\end{itemize}

The first property allows to use galaxy properties (e.g. colour, magnitude, morphology) as an additional information in grouping algorithm\footnote{For example, we can include the information from the red sequence of galaxies as successfully done in redMaPPer cluster finder \citep{2014ApJ...785..104R} for photometric redshift surveys.}. For simplicity, in the current paper we only use the galaxy positions and additional galaxy properties are ignored. This is in line with a widely used FoF clustering algorithm \citep[e.g.][]{2011MNRAS.416.2640R, 2017A&A...602A.100T}, where cluster detection is based only on galaxies positions. The use of galaxy properties as an additional information in grouping algorithm requires specific analysis that will be tackled in future studies.

\begin{figure}
\centering
\includegraphics[width=88mm]{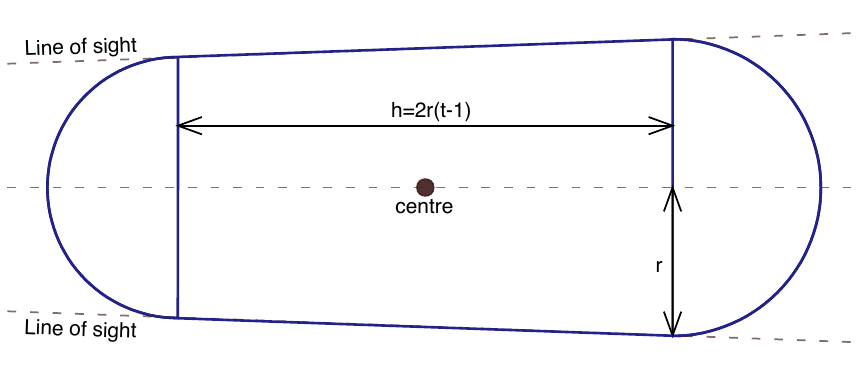}
	\caption{Shape of the object in our marked point process to detect galaxy groups and clusters in cosmological redshift surveys. Figure shows the cross-section of an axisymmetric object. Object consists of two half-balls connected with a truncated cone (a potato shape). Object is fully determined by its centre position, radius $r$ and shape parameter $t \geqslant 1$. Shape parameter $t$ gives the aspect ratio of the object along and perpendicular to the line of sight; for $t=1$ the object is a ball. For a given $r$ and $t$ the height of the truncated cone is defined as $h=2r(t-1)$. The shape of the truncated cone is defined by the lines of sights, which are indicated by dashed lines on the figure. The observer is located at far left from the object.}
	\label{fig:obj_shape}
\end{figure}

The second property allows to consider the cluster pattern as a set of interacting objects, hence using the mathematical framework of marked point processes for describing it. In order to take into account the Finger of God effect, the object generating the cluster pattern has a ``potato'' shape shown in Fig.~\ref{fig:obj_shape}. The shape follows the same idea as used in a FoF algorithm, where the linking length of galaxies along the line of sight is several times larger than the transversal linking length \citep[see e.g.][]{2014MNRAS.440.1763D}. Our objects are elongated along the line of sight that takes into account the local anisotropy of the observed galaxy distribution. The object is fully determined by its centre coordinates, radius $r$ and shape parameter $t \geqslant 1$ that gives the aspect ratio of an object along and perpendicular to the line of sight. The object orientation is determined by the orientation of the line of sight, hence, it is determined by objects centre coordinates. From here on the term object refers to the ``potato'' shape object in our marked point process\footnote{An exception is object point process that we use as an alternative name for marked point process.}.

The object body shown in Fig.~\ref{fig:obj_shape} is a truncated cone with two half-spheres at its extremities, where cone borders follow the lines of sights. This cone shape is only important close to the observer (for small distances) to avoid some pathological cases (i.e. to forbid the connection of galaxies that are not close to each other on the sky plane). In practice, for a majority of the observed volume, the object radius is much smaller than object distance, hence, the object body has nearly cylindrical shape. The knowledge regarding the shape parameters $(r,t)$ is introduced via a probability density. This choice is explained in detail later in the paper.

\subsection{Object point process for cluster pattern detection}
\label{sec:markedpointprocess}

The key hypothesis of our work is that the cluster pattern to be detected is a configuration of random interacting objects driven by the probability density of a marked point process. The solution of the cluster detection problem is given by the construction and manipulation of such a probability density. The probability density we propose takes into account the inhomogeneity of the data, while observational biases and uncertainties in redshift surveys are diminished. Statistical inference using this probability density is done using Markov chains Monte Carlo techniques (see Sections~\ref{sec:simulation} and~\ref{sec:inference}). Such a probability density can be written as
\begin{equation}
	p(\mathbf{y}\,|\,\theta) \propto \exp\left[-U(\mathbf{y}\,|\,\theta)\right],
	\label{eq:gibbs}
\end{equation}
where $U(\mathbf{y}\,|\,\theta)$ is the energy function, $\mathbf{y}$ is the pattern of objects and $\theta$ is the vector of model parameters. The marked point process driven by densities~(\ref{eq:gibbs}) are known in the literature as Gibbs point processes.

The energy function of the cluster pattern can be further written as the sum of two terms:
\begin{equation}
	U(\mathbf{y}\,|\,\theta) = U_\mathcal{D}(\mathbf{y}\,|\,\theta) + U_\mathcal{I}(\mathbf{y}\,|\,\theta),
	\label{eq:utheta}
\end{equation}
where $U_\mathcal{D}$ and $U_\mathcal{I}$ are the data and interaction energy terms, respectively. The data energy term controls the placement of the objects in $W$ depending only on $\mathcal{D}$ so that an object is only placed in a region, where the number density of galaxies (points) is high enough. This term checks the local properties defining the spatial regions of interest. The interaction energy term controls the overlapping of objects so that the unknown spatial regions will be best fitted by object configurations. This term ensures that the overlapping of objects will give the best approximation of the geometrical properties of the galaxy groups and clusters.

The Bayesian framework allows the introduction of the knowledge regarding the parameters via a posterior distribution $p(\theta)$. This allows to write for the joint distribution of the cluster pattern and the model parameters:
\begin{equation}
	p(\mathbf{y},\theta) = p(\mathbf{y}\,|\,\theta)p(\theta).
	\label{eq:probdens}
\end{equation}
Finally, a joint cluster pattern and parameter estimator is given by the maximum of the probability density~(\ref{eq:probdens}):
\begin{equation}
	(\hat{\mathbf{y}},\hat{\theta}) = \mathrm{arg}\,\max_{\Omega\times\Theta}\, p(\mathbf{y},\theta) = 
	\mathrm{arg}\,\max_{\Omega\times\Theta}\, p(\mathbf{y}\,|\,\theta) p(\theta),
\end{equation}
where $\Omega$ is the pattern configuration space and $\Theta$ represents the parameter space. Using the Gibbs energies introduced in Eq.~(\ref{eq:utheta}) we obtain
\begin{equation}
	(\hat{\mathbf{y}},\hat{\theta}) = \mathrm{arg}\,\min_{\Omega\times\Theta}\,
	\left\{ U_\mathcal{D}(\mathbf{y}\,|\,\theta) + U_\mathcal{I}(\mathbf{y}\,|\,\theta) - \mathrm{log}\,p(\theta) \right\} .
	\label{eq:estimator}
\end{equation}
The estimator given by (\ref{eq:estimator}) can be computed using a simulated annealing algorithm \citep{vanLieshout:94,Stoica:05}.

\subsection{Model construction}

Let $W$ be a spatial observation window of volume $\nu(W)$. A simple point process on $W$ is a finite random configuration of points $x_i \in W$, $i=1,\dots,n$ such that $x_i\neq x_j$ whenever $i\neq j$, where $n$ is the number of points in a point process. Characteristics or marks can be attached to the points via a probability distribution. A finite random configuration of marked points is a marked point process if the distribution of the locations only is a simple point process. A point process with marks representing the parameters of geometrical objects, is usually called an object point process. For further reading on marked point processes we recommend the monographs by \citet{vanLieshout:00} and \citet{MollWaag:04}.

The generating object of the cluster pattern is given by $y=(x,r,t)$. The object position is given by $x \in W$. The mark is represented by the shape parameter $(r,t)$ and here, its distribution is the uniform law over $[r_\mathrm{min},r_\mathrm{max}] \times [t_\mathrm{min},t_\mathrm{max}]$ (see Section~\ref{sec:new_methodology} and Fig.~\ref{fig:obj_shape}). Let $\mathbf{y} = \{y_1,y_2,\dots,y_n\}$ be a finite configuration of such objects. The probability density function $p(\mathbf{y}\,|\,\theta)$ controls the positions, marks, and the interaction of objects in the configuration.

The simplest marked point process is the unit rate Poisson point process, where the marks are chosen independently identically according to some distribution. This process does not take into account any interactions between the points. For our problem, more realistic models are constructed by specifying a probability density $p(\mathbf{y}\,|\,\theta)$ that includes interactions between the objects. To specify the interaction energy $U_\mathcal{I}$ in our model, we utilise area-interaction process and pairwise interaction (Strauss like) process~\citep{Strauss:75, Kelly:76, Baddeley:95}. The area-interaction process is able to model the clustering of the objects in a pattern, while the pairwise interaction controls their overlapping. In the following, we describe how we construct the Gibbs energy functions $U_\mathcal{D}$ and $U_\mathcal{I}$ in our model using the framework described above.

\subsubsection{The data energy term $U_\mathcal{D}$}
\label{sec:dataterm}

The data term $U_\mathcal{D}(\mathbf{y}\,|\,\theta)$ is related to the positions of the objects covering the cluster pattern in $W$. For galaxy cluster detection, we want to place the objects in those regions where there are enough galaxies close to each other. Under this consideration, a simple option for the energy contribution of an object $y$ is
\begin{equation}
	v(y)=
	\begin{cases}
		v_\mathrm{const} & \mathrm{if} \quad n_\mathcal{D}(y)\geq n_\mathrm{threshold} \\
		-v_\mathrm{max} & \mathrm{otherwise}
	\end{cases} ,
	\label{eq:energy_object}
\end{equation}
where $v_\mathrm{const}$ and $v_\mathrm{max}$ have positive fixed values and $n_\mathcal{D}(y)$ is the number of points (galaxies) covered by an object $y$. The threshold parameter $n_\mathrm{threshold}$ controls the minimum number of points to be covered by an object. The role of $v_\mathrm{max}$ is to penalise those objects in a configuration that do not fulfil this criterion. If $v_\mathrm{max} \rightarrow \infty$, the configurations with objects not fulfilling the previous condition are forbidden. The threshold value is set $n_\mathrm{threshold}=2$. This choice somewhat mimics the FoF algorithm (see discussion below).

The data energy is the sum of the energy contributions for all the objects in a configuration: 
\begin{equation}
	U_\mathcal{D}(\mathbf{y}\,|\,\theta) = -\sum\limits_{i=1}^{n(\mathbf{y})}v(y_i),
	\label{eq:dataterm}
\end{equation}
where $n(\mathbf{y})$ is the total number of objects in a configuration $\mathbf{y}$. The exponential of the data energy term $\exp\left[-U_\mathcal{D}(\mathbf{y}\,|\,\theta)\right]$ defines the probability density of an inhomogeneous Poisson point process (see Eq.~(\ref{eq:gibbs})). This process is well defined.

We can also make a simple analogy to the commonly used FoF grouping algorithm. We require that the objects used in our model cover at least two galaxies, this can be also seen as linking length in a FoF algorithm. Hence, each object in our model links at least two galaxies. Aside from this similarity, there is an important difference between our model and a FoF algorithm. In our model the size of an object is not fixed (shape parameters are defined using a mark distribution) and it is determined by the model. Additionally, the placement of the objects depends on other objects through interactions (see Section~\ref{sec:interaction_energy}). Most importantly, due to its stochastic nature, compared with the FoF algorithm, our model enables to perform statistical inference related to the detected cluster pattern.

The energy contribution~(\ref{eq:energy_object}) was chosen to be simple on purpose. This term can be written so that to take into account the number density of galaxies or the galaxy properties. In this paper, our aim is to test the point process framework for galaxy group detection and simple energy contribution is sufficient for this purpose. More detailed definition of energy contribution can be considered in future studies.

We note that the data energy term $U_\mathcal{D}$ is the only place in the model, where the observed distribution of galaxies is used. Regarding the application of the proposed methodology to photometric redshift surveys, requires revision of the data energy term, which can also incorporate full photometric redshift posteriors. Hence, the proposed methodology is straightforward to apply to photometric redshift surveys.

\subsubsection{The interaction energy term $U_\mathcal{I}$}
\label{sec:interaction_energy}

If the proposed model would use the data energy term only, some undesirable effects may occur. By construction, the data energy term gives the same probability to objects with different volumes. Another phenomenon that may occur is related to the effect of the maximisation of the probability density of our model (see Section~\ref{sec:markedpointprocess}). The maximisation of a probability density of an inhomogeneous Poisson point process explodes the number of objects in a configuration, while detecting only those cluster regions with the lowest energy function. The role of the interaction energy term is to regularise the solution of our optimisation problem.

The interaction energy term is defined as follows:
\begin{equation}
	U_\mathcal{I}(\mathbf{y}\,|\,\theta) = \nu\left[Z(\mathbf{y})\right]\log\gamma_\mathrm{a} - n_\mathrm{o}(\mathbf{y})\log\gamma_\mathrm{o}.
	\label{eq:Uint}
\end{equation}
The first term of the right-hand side of Eq.~(\ref{eq:Uint}) describes an area-interaction process. The quantity $\nu\left[Z(\mathbf{y})\right]$ represents the Lebesgue measure (volume) of $Z(\mathbf{y})$, the set-theoretic union of all the objects in the configuration $\mathbf{y}$:
\begin{equation}
	Z(\mathbf{y}) = \bigcup_{i=1}^{n(\mathbf{y})} A(y_i)
	\label{eq:intterm_area}
\end{equation}
with $A(y_i)$ the region in observed window $W$ covered by the object $y_i$.

The second term of the right-hand side of Eq.~(\ref{eq:Uint}) describes a pairwise interaction process. The $n_\mathrm{o}(\mathbf{y})$ represents the number of pairs of different objects in $\mathbf{y}$ that overlap:
\begin{equation}
	n_\mathrm{o}(\mathbf{y}) = \frac{1}{2}\sum\limits_{i=1}^{n(\mathbf{y})}\sum\limits_{\substack{ k=1\\ k\neq i }}^{n(\mathbf{y})} \mathbbm{1}\!\left\{ A(y_i)\cap A(y_k) \not= \varnothing \right\},
\end{equation}
where $\mathbbm{1}\!\left\{\cdot\right\}$ is an indicator function that is one if the condition is met and zero otherwise. The model parameters $\log\gamma_\mathrm{a}$ and $\log\gamma_\mathrm{o}$ are specified later by the prior density $p(\theta)$.

Random objects tend to cluster if they are driven by an area-interaction process with a parameter $\gamma_\mathrm{a}>1$. At the same time, such a process helps to fit objects better on the location of galaxy groups and clusters, while reducing the volume occupied by the cluster pattern. Using such a term for the interaction energy implies that a higher probability is given to objects that are more closely clustered together. Simultaneously, smaller objects fit better the data field than big ones, hence reducing the smoothing effects and allowing to adapt the object size individually for each cluster. To prevent the clustering of objects on the best locations of the data term, we use a Strauss-like pairwise interaction process. This process allows objects to superpose, yet, configurations with pairs of objects that overlap are penalised. The repulsion-like interaction prevents the number of objects in a configuration to explode while the maximisation procedure is started. Simultaneously, this interaction forces the objects to spread over the entire spatial domain. The marked point process with probability density $\exp\left[-U_\mathcal{I}(\mathbf{y}\,|\,\theta)\right]$ is well defined. See \citet{Stoica:07} for more details.

\subsection{Simulation method}
\label{sec:simulation}

Several Monte Carlo techniques are available to simulate marked point processes: spatial birth-and-death processes, Metropolis-Hastings (MH) algorithms, reversible jump dynamics or more recent exact simulation techniques \citep{Geyer:94, Green:95, Geyer:99, Kendall:00, vanLieshout:00, vanLieshout:06}.

In this paper, we need to sample from the joint law $p(\mathbf{y}\,|\,\theta)$. This is done by using an iterative Monte Carlo algorithm, where an iteration consists of two steps. First, a parameter value is chosen with respect to the prior law of the model parameters $p(\theta)$. Then, conditionally on $\theta$, an object pattern is sampled from $p(\mathbf{y}\,|\,\theta)$ using an MH algorithm \citep{Geyer:94, Geyer:99}. The MH algorithm consists of three types of moves.

(\emph{i}) Birth: with a probability $p_\mathrm{b}$ a new object $\zeta$, sampled from the birth rate $b(\mathbf{y},\zeta)$, is proposed to be added to the present configuration $\mathbf{y}$. The new configuration $\mathbf{y}\prime = \mathbf{y}\cup\zeta$ is accepted with the probability
\begin{equation}
	\min\left\{ 1,\frac{p_\mathrm{d}}{p_\mathrm{b}}
	\frac{d(\mathbf{y}\cup\zeta,\zeta)}{b(\mathbf{y},\zeta)}
	\frac{p(\mathbf{y}\cup\zeta)}{p(\mathbf{y})} \right\}.
	\label{eq:pbirth}
\end{equation}

(\emph{ii}) Death: with a probability $p_\mathrm{d}$ an object $\zeta$ from the current configuration $\mathbf{y}$ is proposed to be eliminated according to the death proposal $d(\mathbf{y},\zeta)$. The probability of accepting the new configuration $\mathbf{y}\backslash \zeta$ is computed reversing the ratio~(\ref{eq:pbirth}).

(\emph{iii}) Change: with a probability $p_\mathrm{c}$ we randomly choose an object $\zeta_\mathrm{old}$ in the configuration $\mathbf{y}$ and propose to slightly change its parameters using uniform proposals. For the selected object, we may change its location within the vicinity $\Delta k$ of its centre and change its shape parameters $(r,t)$ within a small tolerance with respect to its initial values. The new object obtained is $\zeta_\mathrm{new}$. The new configuration $\mathbf{y}\prime = \mathbf{y}\setminus\zeta_\mathrm{old}\cup\zeta_\mathrm{new}$ is accepted with the probability $\min\{ 1, p(\mathbf{y}\prime)/p(\mathbf{y}) \}$.

For the birth and death rates, we adopt the uniform choices $b(\mathbf{y},\zeta) = 1/\nu(W)$ and $d(\mathbf{y},\zeta) = 1/n(\mathbf{y})$, where $\nu(W)$ is the Lebesgue measure (volume) of the observed window $W$ and $n(\mathbf{y})$ is the number of objects in the configuration.

In order to maximise $p(\mathbf{y},\theta)$, the previously described sampling mechanism is integrated into a simulated annealing algorithm. The simulated annealing algorithm is built by sampling from $p(\mathbf{y},\theta)^{1/T}$, while $T$ goes slowly to zero. The authors in \citet{Stoica:05} proved the convergence of a simulated annealing algorithm for marked point processes, if MH dynamics and a logarithmic cooling schedule are used. According to this result, the temperature is lowered as
\begin{equation}
	T_k = \frac{T_0}{\mathrm{log}\,k + 1},
\end{equation}
where $T_0$ is the initial temperature and $k$ is a time-step in a simulation.

\subsection{Inference}
\label{sec:inference}

The problem we are trying to solve is the cluster pattern detection. In the framework of our model, the question we are interested in is what are the centre positions and shape parameters of the objects approximating the spatial regions induced by the cluster pattern exhibited by the point field $\mathcal{D}$. The only thing we can observe here is the data point field $\mathcal{D}$. The objects approximating the spatial regions induced by the cluster pattern and the model parameters are unknown. 

The optimal objects configurations are estimated using the estimator given by (\ref{eq:estimator}), which is computed using MH and simulated annealing algorithms (see Section~\ref{sec:simulation}). Additionally, we have to choose a prior law $p(\theta)$ for model parameters. Since prior knowledge about model parameters is not available, we use fixed intervals for the model parameters. The values of the intervals are chosen based on previous knowledge and a trial and error analysis. Section~\ref{sec:application} gives the model parameter values and ranges that we used for the 2MRS data set.

Due to the complexity of cluster pattern detection, the optimisation algorithm does not provide a unique solution. Hence, a question arises: what is the probability that a given region in the spatial domain (or a given galaxy) belongs to the cluster pattern? Our proposed method use level sets estimation to answer this question \citep{heinrich:12}.

The probability that a given region $\mathcal{R}\subset W$ is covered by the cluster pattern configuration is
\begin{equation}
	p_\mathrm{visit}(\mathcal{R}) = 
	\mathbbm{P}\left[ \mathcal{R} \subseteq \mathbf{Y} \right]
	= \mathbbm{E}\left[ \mathbbm{1}\!\left\{ \mathcal{R} \subseteq \mathbf{Y} \right\} \right].
	\label{eq:pvisit}
\end{equation}
The $p_\mathrm{visit}$ represents the visiting probability of region $\mathcal{R}$ by the clustered pattern in one configuration. The $p_\mathrm{visit}$ can be estimated as follows
\begin{equation}
	\hat{p}_\mathrm{visit}(\mathcal{R}) = \frac{1}{n(\mathbf{Y})} \sum\limits_{i=1}^{n(\mathbf{Y})} \mathbbm{1}\!\left\{ \mathcal{R} \subseteq \mathbf{Y}_i \right\},
\end{equation}
where $\mathbf{Y}_i$, $i=1,\dots,n(\mathbf{Y})$ are cluster patterns simulated with our model.

If $\mathcal{R} = \{x\}$ with $x \in W$ then $p_\mathrm{visit}(\mathcal{R})$ is the probability that the cluster pattern touches the point $x$. Under a stationary hypothesis, which is not fulfilled here, this quantity may represent the volumic fraction of the clustered pattern~\citep{Chiu:13}. The computation of this probability in every point of $W$ gives a probability field, that we call the visit map. Peaks in the visit map may correspond to individual galaxy groups and clusters.

If $\mathcal{R} = \{x,y\}$ with $x,y \in W$ then $p_\mathrm{visit}(\mathcal{R})$ is the probability that the cluster pattern touches simultaneously the points $x$ and $y$. This is known also as the covariance function~\citep{Chiu:13}. Under a similar stationary hypothesis as previously, the covariance function depends only on the distance between the points $x$ and $y$. This probability may be used to compute the probability that two galaxies belong to the same cluster. The quantities $p_\mathrm{visit}({x})$ and $p_\mathrm{visit}({x,y})$ are used to extract individual galaxy clusters and to determine to which cluster a galaxy belongs to. All this is described in Section~\ref{sec:group_extraction}.

\subsection{Special considerations for flux-limited galaxy surveys}
\label{sec:fluxlim_consideration}

In flux-limited galaxy surveys, the number density of galaxies decreases as a function of distance from the observer. It means that the data field $\mathcal{D}$ is inhomogeneous (it depends on distance). If we do not take this into account in our model then it means that the probability to have two galaxies in a given object decreases as a function of distance from the observer. This is an undesirable effect. The decrease of number density of galaxies can be compensated while increasing the size of objects as a function of distance from the observer. Similar approach is used in FoF grouping algorithms, where the linking length increases as a function of distance from the observer \citep[e.g.][]{2017A&A...602A.100T}.

To compensate the inhomogeneity of the data field $\mathcal{D}$, the object radius at a distance $d$ is defined as
\begin{equation}
	r(d) = \left[ r_0(d)f_{r,\mathrm{min}}, r_0(d)f_{r,\mathrm{max}} \right],
	\label{eq:radius}
\end{equation}
where $r_0(d)$ is a distance dependent function that describes how object radius depends on distance, and $f_{r,\mathrm{min}}<1.0$ and $f_{r,\mathrm{max}}>1.0$ are some pre-defined constants. At a fixed distance $d$ the object radius has a uniform law over $[r_0(d)f_{r,\mathrm{min}}, r_0(d)f_{r,\mathrm{max}}]$. The function $r_0(d)$ is an analytical function that is determined based on the observed data as described in Section~\ref{sec:2mrs_settings}. For the object shape $t$ (see Fig.~\ref{fig:obj_shape}) we use uniform law over $[t_\mathrm{min},t_\mathrm{max}]$ that is independent of the distance from the observer. In this case the distance from the observer $d$ and the mean object volume $\overline{V}_y$ at distance $d$ are related as
\begin{equation}
	\overline{V}_y(d) \propto \left[r_0(d)\right]^3 .
	\label{eq:vol_prop_to_dist}
\end{equation}

The change of object radius as a function of distance (Eg.~(\ref{eq:radius})) compensates the decrease of the number density of galaxies. The aim of this compensation is to ensure that the data energy given by Eq.~(\ref{eq:energy_object}) is roughly independent of the distance from the observer. This can be achieved by choosing an adequate form for the function $r_0(d)$ (see Section~\ref{sec:2mrs_settings}). In practice, it means that the local definition of a cluster changes as a function of distance from the observer, but since the data energy term is independent of distance, the general cluster pattern is left intact as much as possible.

In our point process model we have two terms, data energy term and interaction energy term. The interaction energy term depends on the volume of the object (see Eq.~(\ref{eq:intterm_area})), the probability to add an object to the configuration is proportional to the volume of the object (see Section~\ref{sec:interaction_energy}). While introducing a distance dependent radius for an object, the object volume also depends on distance from the observer as given by Eq.~(\ref{eq:vol_prop_to_dist}). This means that the probability to add a new object decreases as a function of distance from the observer (since the mean object volume increases as a function of distance), which is an undesirable side effect. To compensate this, a volume element $\mathrm{d}V$ at distance $d$ is multiplied by $\left[r_0(d)\right]^{-3}$. This can be seen as squeezing of the space so that the mean object volume is independent of the distance. In our point process model it affects the area interaction part in the interaction energy term (see Eq.~(\ref{eq:Uint})). The modification of the volume element can be visualised as following. If the area that all objects cover are calculated using a regular grid, then the volume of each grid cell is multiplied by $\left[r_0(d_\mathrm{cell})\right]^{-3}$, where the $d_\mathrm{cell}$ is a cell distance from the observer.

This introduced modification of object radius (and volume) ensures that the probability to add a new object to the configuration does not depend strongly on distance from the observer, despite the fact that the galaxies distribution depends on it. The used point process model for cluster pattern detection assumes homogeneity of the data field $D$, which is violated in flux-limited galaxy redshift surveys. The distance dependent object radius and the corresponding squeezing of space (volume elements) are introduced to tackle this inhomogeneity. This is a compromised solution, it requires more detailed studies to find an alternative approach.

\section{Extracting galaxy groups and assigning galaxies memberships}
\label{sec:group_extraction}

This section presents how the quantities $p_\mathrm{visit}({x})$ and $p_\mathrm{visit}({x,y})$, defined with Eq.~(\ref{eq:pvisit}), are used for the galaxies classification into groups.
The extraction of galaxy groups and clusters (and the construction of group catalogue) is complicated because the number of groups and clusters is not known in advance and even for relatively isolated systems the boundaries and membership of groups and clusters is fuzzy. The situation is even more complicated for merging and/or nearby systems. Hence, it is possible to construct several algorithms for group extraction that all yield to slightly different results. Below we describe just one such algorithm based on hierarchical clustering that is simple enough and, most importantly, parameter free. It deserves a special study, whether other algorithms, for example fuzzy k-means \citep{Bezdek:81} or neural network based methods \citep{Ripley:96}, can be meaningfully applied to our marked point process output. Most likely, depending on the goal of the specific study, different algorithms can be considered as the preferred ones.

Our group extraction algorithm is done iteratively consisting of two steps. Before we start these two steps, we calculate $p_\mathrm{visit}({x,y})$ between all galaxy pairs. To suppress small computational noise in our marked point process, we apply a minimum threshold value for covariance function
\begin{equation}
	p_\mathrm{visit}({x,y}) = 
	\begin{cases}
		p_\mathrm{visit}({x,y}) & \mathrm{if} \quad p_\mathrm{visit}({x,y}) > c_\mathrm{noise\_limit} \\
		0 & \mathrm{otherwise}
	\end{cases},
\end{equation}
where $c_\mathrm{noise\_limit}$ is set to be few per cent from the maximum covariance function value.

\begin{figure}
\centering
\includegraphics[width=88mm]{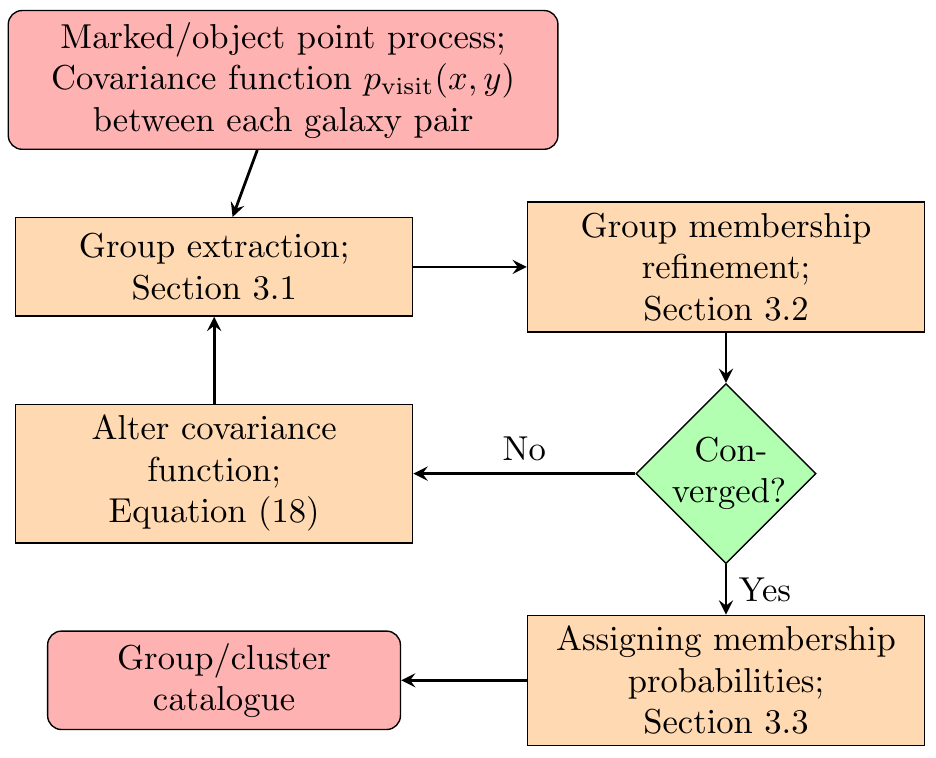}
	\caption{Block diagram outlining the group extraction and catalogue generation procedure. Input for the group extraction is the covariance function $p_\mathrm{visit}({x,y})$ between each galaxy pairs. Group extraction procedure converges if group membership refinement does not expel any group members. The output of group extraction procedure is a catalogue of groups and clusters that is described in Appendix~\ref{app:cat}. See Section~\ref{sec:group_extraction} for more details.}
	\label{fig:group_making}
\end{figure}

In the first step of group extraction, we use the calculated covariance function $p_\mathrm{visit}({x,y})$ values and assign each galaxy to a single group using hierarchical clustering as described in Section~\ref{sec:group_extraction_algorithm}. In the second step, we use the group refinement procedure and expel all physically unbound galaxies from the groups as described in Section~\ref{sec:group_refinement}. After refinement procedure, we adjust the covariance functions for expelled galaxies and go back to the step one and repeat the procedure until there are no expelled galaxies. This iterative procedure is outlined in Fig.~\ref{fig:group_making}. In case of 2MRS data set, it takes around 30 iterations before it converges.

The adjustment of the covariance function for expelled galaxies is done in the following way. Let $x_\mathrm{G,expel}$ be an expelled galaxy from group $G$, which after refinement contains galaxies $y_\mathrm{G,1},y_\mathrm{G,2},\dots,y_{\mathrm{G},n(G)}$, where $n(G)$ is the number of galaxies in group $G$ after the refinement procedure. Hence, before refinement galaxy $x_\mathrm{G,expel}$ belongs to the group $G$, but after refinement it does not. For every expelled galaxy $x_\mathrm{G,expel}$, the covariance function with the group members it used to belong to is set to zero,
\begin{equation}
	p_\mathrm{visit}(x_\mathrm{G,expel},y_{\mathrm{G},i}) = 0, \quad \forall i \in\{1,2,\dots,n(G)\}.
\end{equation}
This step ensures that physically unbound galaxies are not considered as group members in the subsequent iterations.

After iterative group extraction is converged, we assign to each galaxy a probability that it belongs to a given group or cluster. One galaxy can belong to several groups if the groups are nearby or merging systems. The probability assignment is based on the initial covariance function $p_\mathrm{visit}({x,y})$ and is described in Section~\ref{sec:member_probabilities}. The membership probabilities together with group main properties (e.g. location, size, mass) are made available in our group catalogue (see Appendix~\ref{app:cat}).

\subsection{Group extraction algorithm}
\label{sec:group_extraction_algorithm}

The aim of group extraction procedure is to group together galaxies that form a single system (a group or cluster). Group extraction in this paper is based on an agglomerative hierarchical clustering, where a measure of ``distance'' between galaxy $x_i$ and group $G_j$ is defined as
\begin{equation}
	d(x_i,G_j) = \sum\limits_{y_k\in G_j, \, y_k\neq x_i} p_\mathrm{visit}(x_i,y_k),
	\label{eq:distance_measure}
\end{equation}
where the summation is over galaxies $y_k$ that belong to the group $G_j$. The distance measure (\ref{eq:distance_measure}) allows to find the closest group to each galaxy $x_i$ as follows
\begin{equation}
	G_{x_i,\mathrm{closest}} = \operatorname*{arg\,max}_{ G_j\in[G_1,\dots,G_{N_G}] } \, d(x_i,G_j),
\end{equation}
where $N_G$ is the total number of groups.

To start the hierarchical clustering, initially each galaxy is assigned to its own group, that means the number of galaxies and groups is the same. During one step down in hierarchical clustering, we merge groups $G_i$ and $G_j$ if the following criterium is satisfied
\begin{equation}
	\exists y_k\in G_i: G_{y_k,\mathrm{closest}} = G_j.
	\label{eq:merge_crit}
\end{equation}
Using criterium (\ref{eq:merge_crit}) we test all possible group pairs $G_i$ and $G_j$ and merge all of them that satisfy the criterium.

In hierarchical clustering we go down only two steps and then stop the algorithm. Most of the formed groups after first step contain only two galaxies, these are galaxy pairs that are most strongly connected to each other. During the second step down, isolated systems remain isolated but groups that are relatively close to each other are merged together.

The used algorithm for clustering is essentially a parameter free algorithm (the only parameter is the noise limit $c_\mathrm{noise\_limit}$) and depends only on the probabilisitic output of our model ($p_\mathrm{visit}$). In general, more advanced methods can be used for group extraction. In this paper we prefer to use a simple algorithm in order to test the marked point process framework for group detection in cosmological redshift surveys, without altering it with sophisticated post-processing techniques.

\subsection{Group membership refinement procedure}
\label{sec:group_refinement}

The aim of group membership refinement is to expel physically (gravitationally) unbound members from extracted systems. This procedure is purely based on a physical considerations and it is not connected with the point process framework. For membership refinement, each group is analysed separately using the extracted group members (see Section~\ref{sec:group_extraction_algorithm}). The membership refinement procedure follows directly the procedure described in \citet{2016A&A...588A..14T}. Below we provide a brief summary of it.

Group membership refinement is based on estimates of the virial radius and escape velocity of the system. Galaxy is expelled from its group if its projected distance from the group centre in the plane of the sky is greater than the virial radius of the system. Similarly, a galaxy is removed from its group if the velocity of the galaxy with respect to the group centre is higher than the escape velocity at its sky-projected distance from the group centre. The escape velocity of a group relates to the gravitational potential $\Phi$ through
\begin{equation}
	v^2_\mathrm{esc}(r) = -2\Phi(r),
	\label{eq:escvel}
\end{equation}
where $r$ is projected distance from the group centre.

To calculate the virial radius and escape velocity of a system we have to assume some dark matter density profile. To calculate these quantities, we follow \citet{2014A&A...566A...1T}. Group mass is estimated using the virial theorem and assuming NFW profile \citep{1997ApJ...490..493N}
\begin{equation}
	M_\mathrm{vir} = 2.325 \frac{R_\mathrm{g}}{\mathrm{Mpc}}\left(\frac{\sigma_v}{100\,\mathrm{km}\,\mathrm{s}^{-1}}\right)^2 10^{12}M_\odot,
	\label{eq:mass}
\end{equation}
where $R_\mathrm{g}$ is the gravitational radius, which for a fixed mass density profile only depends on the group extent in the sky $\sigma_\mathrm{sky}$. See \citet{2014A&A...566A...1T} for details about mass and gravitational radius calculations. Under the assumption of an NFW profile, the group virial radius is uniquely defined with the virial mass, which radius is defined as the radius in which the mean density is 200 times higher than the mean density of the Universe. Gravitational potential, needed to calculate the escape velocity in Eq.~(\ref{eq:escvel}), is directly related to the assumed dark matter density profile \citep[see e.g.][]{2001MNRAS.321..155L}.

In practice, the group membership refinement is purely defined by the group velocity dispersion $\sigma_v$ and the group extent in the sky plane $\sigma_\mathrm{sky}$. These quantities are defined with the following formulas:
\begin{equation}
	\sigma_v^2 = \frac{1}{2n(1+z_\mathrm{m})}\sum\limits_{i=1}^{n}(r_i)^2
\end{equation}
and
\begin{equation}
	\sigma_\mathrm{sky}^2 = \frac{1}{(1+z_\mathrm{m})(n-1)}\sum\limits_{i=1}^{n}(v_i-v_\mathrm{m})^2,
\end{equation}
where $z_\mathrm{m}$ and $v_\mathrm{m}$ are the mean redshift and velocity of the group; $v_i$ and $r_i$ are velocities and the projected distances for individual group members. Summation is over all galaxies within the group.

We note that the approach for membership refinement is somewhat conservative, the sky-projected distance generally underestimates the three dimensional distance, thus we tend to overestimate the escape velocity, leaving some outliers in the group rather than removing true group members.

The membership refinement procedure is done iteratively since the group velocity dispersion and size depend on the group membership. During the iteration process, initially the virial radius and escape velocity was multiplied by a factor of ten. The multiplication factor was gradually lowered during each iteration until it reaches unity. This ensures that during one iteration only few galaxies were excluded and the refinement procedure converges.

\subsection{Assigning probabilities to group members}
\label{sec:member_probabilities}

In Sections~\ref{sec:group_extraction_algorithm} and~\ref{sec:group_refinement} we described how we extract galaxy groups and assign galaxies to individual systems. Our probabilistic framework for group detection allows to do more than just assigning galaxies to individual systems. Using the covariance function $p_\mathrm{visit}({x,y})$ we can compute how strongly each galaxy is connected with any group. This allows us to define for each galaxy a probability that it belongs to any detected system. The probability that galaxy $x_i$ belongs to detected group $G_j$ is calculated as
\begin{equation}
	p(x_i,G_j) = \frac{1}{C_i} \sum\limits_{y_k\in G_j, \, y_k\neq x_i} p_\mathrm{visit}(x_i,y_k)\,
	\mathbbm{1}\!\left\{\mathrm{Ref.:}\, x_i \in G_j \right\} ,
	\label{eq:galprob}
\end{equation}
where $C_i$ is the normalizing constant for galaxy $x_i$ defined as
\begin{equation}
	\sum\limits_{j=1}^{N_G} p(x_i,G_j) = 1.0
\end{equation}
and $N_G$ is the total number of detected (extracted) galaxy groups. The probabilities defined by Eq.~(\ref{eq:galprob}) include the constraints from group refinement procedure. The notation $\mathbbm{1}\!\left\{\mathrm{Ref.:}\, x_i \in G_j \right\}$ is one if galaxy $x_i$ satisfies the group refinement criteria defined in Section~\ref{sec:group_refinement} and is zero otherwise. The group properties are not altered while evaluating the group refinement criteria. The idea of this requirement is to avoid assigning galaxies to groups that are not physically (gravitationally) bound to these systems.

For majority of the galaxies, the probability $p(x_i,G_j)$ is non-zero (exactly one) only for one galaxy group, that means a galaxy belongs to only one group or the galaxy is an isolated (does not belong to any systems). In case of nearby and/or merging systems, galaxies in the outer parts of groups can belong to several systems. For each galaxy in our catalogue, we calculate the probability that it belongs to any of the systems and provide this information in our group catalogue.

\section{Application of the model to the 2MRS data set}
\label{sec:application}

In this section we apply the proposed methodology to the 2MRS data set. The main aim is to test the feasibility of the Bayesian methodology using a relatively small observational data set. 2MRS is well suited for this task since there are well studied FoF group catalogues available \citep[e.g.][]{2016A&A...588A..14T}.
In Appendix~\ref{app:mock} we apply our method to a simulated mock catalogue and analyse how the proposed method recovers the true members of groups and clusters.

\subsection{Galaxies in 2MRS data set}

The proposed model for galaxy group detection is applied to the 2MRS data set described in \citet{2012ApJS..199...26H}. This data set includes galaxies brighter than 11.75~mag in the $K_\mathrm{S}$ band. The galaxy sample is downloaded from the extragalactic distance database (EDD\footnote{\url{http://edd.ifa.hawaii.edu}.}; \citealt{2009AJ....138..323T}).

The 2MRS data set is highly complete above the Galactic plane, Galactic latitude $|b|>5\degr$. The completeness is slightly lower in very nearby Universe due to the low surface brightness galaxies \citep[see][]{2013AJ....145..101K}. 2MRS galaxy sample becomes very sparse father away, hence we restricted ourselves with galaxies closer than 300~Mpc. According to the FoF group catalogue constructed by \citet{2016A&A...588A..14T} this includes all galaxy groups of at least five members. This selection restricts our 2MRS sample to 42620 galaxies.

\subsection{Setting model parameters for group detection}
\label{sec:2mrs_settings}

To apply the proposed cluster pattern detection model to the 2MRS data set we have to fix some of the parameters in the model. There are two types of parameters. The first set contains parameters related to the model, the second set contains parameters related to the simulation of the model that mostly affect the efficiency and convergence of the point process simulation.

The parameters that most strongly affect the group detection are the shape parameters of the object in the marked point process. The shape parameters are the radius $r$ and ratio $t$ (see Section~\ref{sec:new_methodology} and Fig.~\ref{fig:obj_shape}). We chose the values of these parameters following the linking length values used to construct the FoF groups in \citet{2016A&A...588A..14T}. The ratio is chosen to be $t\in [5,12]$ with uniform law in this range. Since the object in our model is a linking length analogue in a FoF algorithm, the chosen parameter range for the ratio $t$ is well justified \citep[see][]{2014MNRAS.440.1763D}.

\begin{figure}
\centering
\includegraphics[width=88mm]{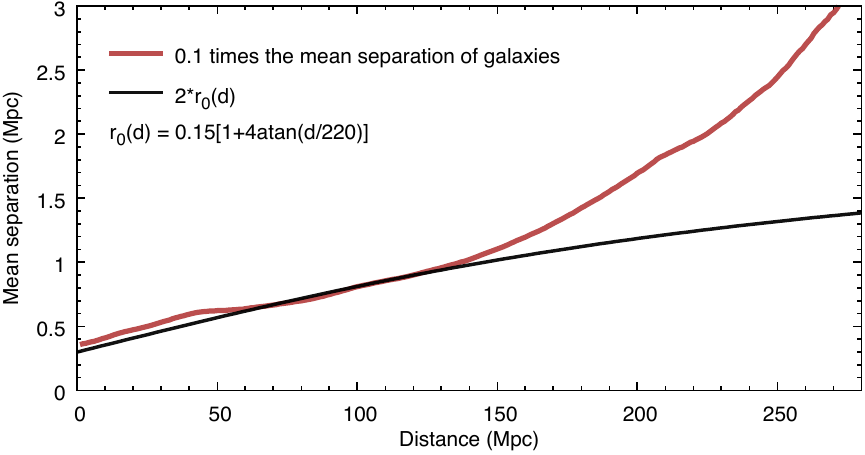}
	\caption{Mean separation between the galaxies as a function of distance from the observer in the 2MRS data set. Red line shows the mean nearest neighbour separation between galaxies (multiplied by $0.1$), which is usually taken as the linking length in a FoF algorithm. Black line shows the arctan function that we use to describe the base object radius in our marked point process.}
	\label{fig:2mrs_objects}
\end{figure}

As noted in Section~\ref{sec:fluxlim_consideration} in flux-limited surveys the number density of galaxies decreases as a function of distance. To take this into account, the object radius $r$ depends on the distance from the observer. We follow the considerations from FoF algorithm, where the linking length in transversal direction is approximately $0.1$ times the mean distance between galaxies in the survey \citep[see e.g.][]{2014A&A...566A...1T}. In Fig.~\ref{fig:2mrs_objects} we show the mean distance between galaxies in 2MRS as a function of distance from the observer. Farther away from the observer, the mean separation increases rapidly because most of the galaxies at these distances are the central galaxies of individual groups or clusters. To detect groups, where groups parameters are roughly constant as a function of distance, in a FoF algorithm a moderately increasing linking length value is used \citep[see][]{2014A&A...566A...1T, 2016A&A...588A..14T}. We follow the same approach here and increase the object base radius $r_0$ according to the arctan function
\begin{equation}
	r_0(d) = 0.15\left[1+4\,\mathrm{atan}(d/220)\right],
\end{equation}
where $d$ is distance from the observer in Mpc. This scaling follows the $0.1$ times mean distance between galaxies in nearby regions and the increase at farther distances is rather shallow (see Fig.~\ref{fig:2mrs_objects}). In our model we use a range for object radius $r$ given with Eq.~(\ref{eq:radius}), where we fix $f_{r,\mathrm{min}}=0.5$ and $f_{r,\mathrm{max}}=1.5$. The chosen parameter ranges for $r$ and $t$ are sufficiently wide and provide reasonable mark distribution for our point process model. The ranges for model parameter values are chosen using similar considerations as in a FoF algorithm, which takes into account the peculiarities of the observed data set. However, the shape parameters in our model are not fixed and the best values at each location are determined in a Bayesian manner.

The remaining parameters that we have to specify are related to the point process model itself. The most important parameters are the interaction parameters where we use the following ranges: $\log\gamma_\mathrm{a}=[0.12,0.18]$ and $\log\gamma_\mathrm{o}=[-0.1,-0.06]$. We have no prior knowledge about the interaction parameters, so we have chosen the uniform distribution over these parameters. We fix the constant data energy parameter $\nu_\mathrm{const}=0.36$, which equals to the mean interaction energy of added objects. The last quantity and the interaction parameters are not priorly known and therefore they were estimated based on trial runs of the simulation. In general, these parameters affect mostly the efficiency of the cluster pattern detection (how well clusters are detected in the data set) and have minor effect on the physical parameters of the individual detected systems. For future applications, we will consider the ABC Shadow algorithm \citep{Stoica:17} to estimate the model parameters in an automated fashion.

The object point process simulation is carried out using the MH algorithm with simulated annealing (see Section~\ref{sec:simulation}). The simulation parameters were set following the same principles as we used in the Bisous model for filamentary pattern detection \citep{2014MNRAS.438.3465T, 2016A&C....16...17T}. The initial temperature in simulated annealing was set to $T_0=1.2$. In our MH simulation, one move concerns only one object. The temperature was lowered after 30\,000 moves using the logarithmic cooling schedule and the final temperature in our simulation was around $0.1$.

To make an inference from the model as described in Section~\ref{sec:inference}, we extracted several realisations of the objects configurations. To extract realisations that are statistically uncorrelated, there should be a sufficient number of moves between realisations. For our analysis, we extracted realisations after 1.2 million moves. We run 8 simulations (the same model parameters but different random seed), and from each simulation we extracted 2500 realisations. For the inference, we only use last 1500 realisations, which gives us in total 12\,000 realisations of the objects configurations for visit map and covariance function calculations.

\subsection{Remarks on the practical implementation of the model}

	Regarding the practical implementation of the model, there are several numerical difficulties. The proposed model requires the calculation of the total volume covered by individual small objects. Since the geometry of the objects is complicated and most of the objects are overlapping, this volume cannot be calculated analytically. To calculate the volume covered by objects, we use numerical approach. The total volume is divided into small cells with a cell size of $(0.26~\mathrm{Mpc})^3$. To calculate the total volume covered by objects, we sum the cells that are covered by small objects (cell centre is inside the object). To mitigate the effect of pre-defined grid pattern, the grid origin (first grid point) is slightly shifted after a certain number of iterations (after one realisation of objects configuration is extracted). Another numerical difficulty is the counting of overlapping objects. This is solved using the same grid cells that we use for the volume calculation. If two objects cover the same grid cell, then these objects are considered overlapping.

	In order to calculate the statistics described in Section~\ref{sec:inference} we need many realisations. In our 2MRS volume, there are approximately 7000 groups and clusters. In our marked point process, these groups are detected using roughly 30\,000 objects. Hence, one realisation of our model consists of approximately 30\,000 objects. In order to calculate the statistics, we need thousands of realisations. In order to achieve this, we initially store all realisations (objects locations and shapes) on hard drive and use a post processing of the realisations to calculate the necessary statistics.
	
	The use of the proposed object point process model is computationally expensive. In case of 2MRS data set, to run a single simulation (in one core) it took around 10 days using a personal computer (3~GHz and 32~GB of RAM). For the future applications, we will investigate how to optimise the modelling procedure in order to apply it to larger data sets using parallel computations.

\section{Galaxy groups in 2MRS data set}
\label{sec:2mrs_groups}

\subsection{General properties and selected examples: Coma and Virgo clusters}

We extracted the galaxy groups from the 2MRS data set as described in Section~\ref{sec:application}. Our group catalogue contains 7755 systems (groups and clusters) with at least two members and 1933 systems with at least three galaxies. Hence, most of the detected systems in our catalogue are galaxy pairs. The catalogue of 2MRS groups includes 95 systems with more than ten members. We visually looked at all these 95 systems and verified that they are reasonable galaxy systems and that our Bayesian grouping algorithm works as expected. 

\begin{figure}
\centering
\includegraphics[width=88mm]{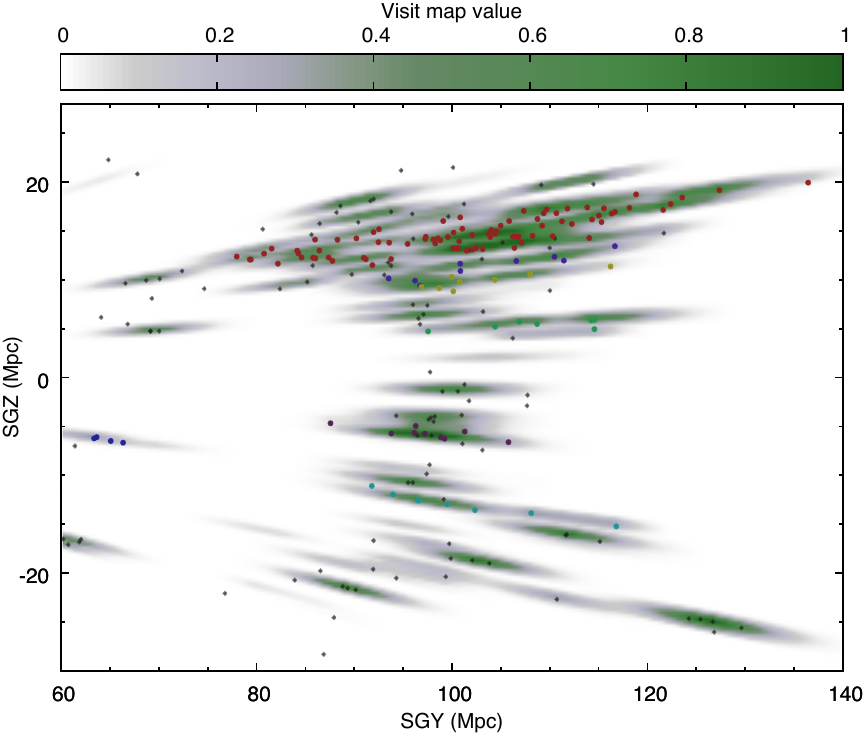}
	\caption{The distribution of galaxies in supergalactic coordinates (points) and the visit map detected using our marked point process. The thickness of the slice is 4~Mpc around $\mathrm{SGX}=0$. Red points show galaxies in the Coma cluster (see also Fig.~\ref{fig:coma}), other coloured points show galaxies in other groups with at least five members, and grey points show all remaining galaxies. Galaxies are divided into groups as explained in Section~\ref{sec:group_extraction}.}
	\label{fig:visitmap}
\end{figure}

Figure~\ref{fig:visitmap} illustrates the detection of galaxy groups and clusters in our model. The green colour in Fig.~\ref{fig:visitmap} shows the visit map in a region of the \object{Coma cluster}. For isolated systems, the visit map shows clear maxima and individual groups are clearly distinguishable. In the Coma cluster region (red points in the figure), the situation is much more complicated showing a rich inner structure (see discussions below). However, the entire Coma cluster region is connected through the visit map.

\begin{figure}
\centering
\includegraphics[width=88mm]{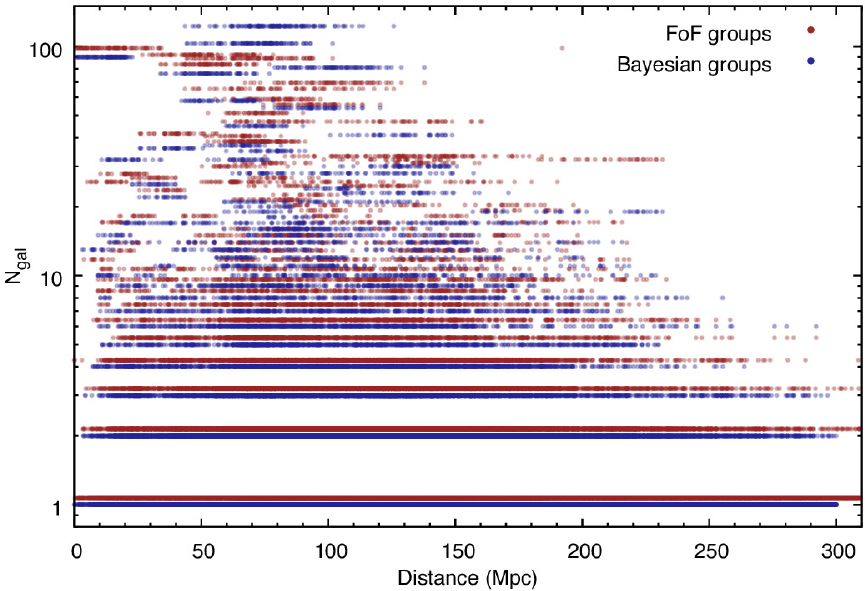}
	\caption{Number of galaxies in a group ($N_\mathrm{gal}$) as a function of distance from the observer. We plot all galaxies in groups where the distance is a comoving distance calculated using the heliocentric velocities of galaxies. Blue points show the number of galaxies in Bayesian groups as extracted in this study. Red points show the number of galaxies in the FoF groups as constructed by \citet{2016A&A...588A..14T}. Red points are shifted slightly upward to increase the readability of the figure.}
	\label{fig:ngal_vs_dist}
\end{figure}

In Fig.~\ref{fig:ngal_vs_dist} we show the number of galaxies in clusters as a function of distance from us. Since we plot all galaxies in every group, the rich groups appear as horizontal lines of points (due to the peculiar motions of galaxies inside the groups). We limited ourselves with galaxies closer than 300~Mpc in comoving distances. From Fig.~\ref{fig:ngal_vs_dist} we can see that this includes all systems with at least five members. The galaxy systems farther away than 300~Mpc are very poor systems due to the flux-limited nature of the 2MRS data set. As a comparison, in Fig.~\ref{fig:ngal_vs_dist} we also show the group richness in the FoF groups detected in \citet{2016A&A...588A..14T}. The general distribution of group richnesses as a function of distance is the same in both catalogues. In general, new Bayesian group catalogue is very similar with the FoF group catalogue published by \citet{2016A&A...588A..14T}. More detailed comparison between these two catalogues are carried out in Section~\ref{sec:comparison}.

\begin{figure*}
\centering
\includegraphics[width=182mm]{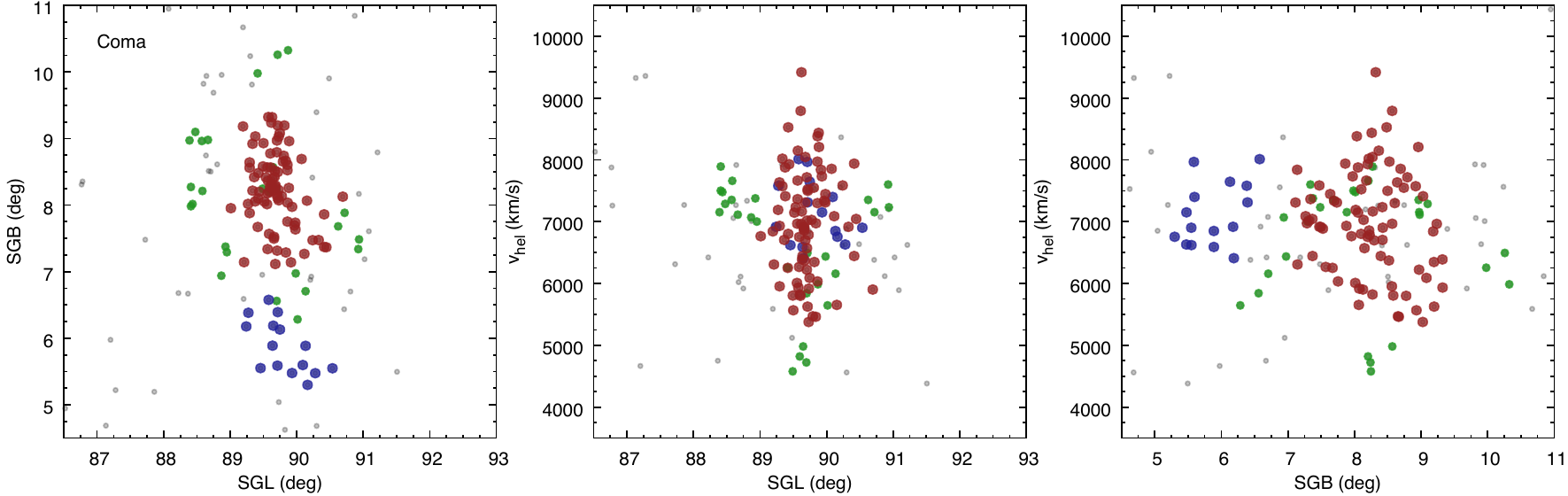}
	\caption{Coma cluster region in 2MRS data set. Left panel shows the galaxies in supergalactic longitude (SGL) and latitude (SGB) between heliocentric velocities 3500 and 10\,500~$\mathrm{km}\,\mathrm{s}^{-1}$. Middle and right panels show the galaxies in supergalactic coordinates and heliocentric velocity plane. Red points show the central component of the Coma cluster detected using the Bayesian group finder. Blue points show another relatively large component (15 members) and green points show the galaxies in groups with three to four members. Grey points show all other galaxies in Coma cluster region.}
	\label{fig:coma}
\end{figure*}

\begin{figure*}
\centering
\includegraphics[width=182mm]{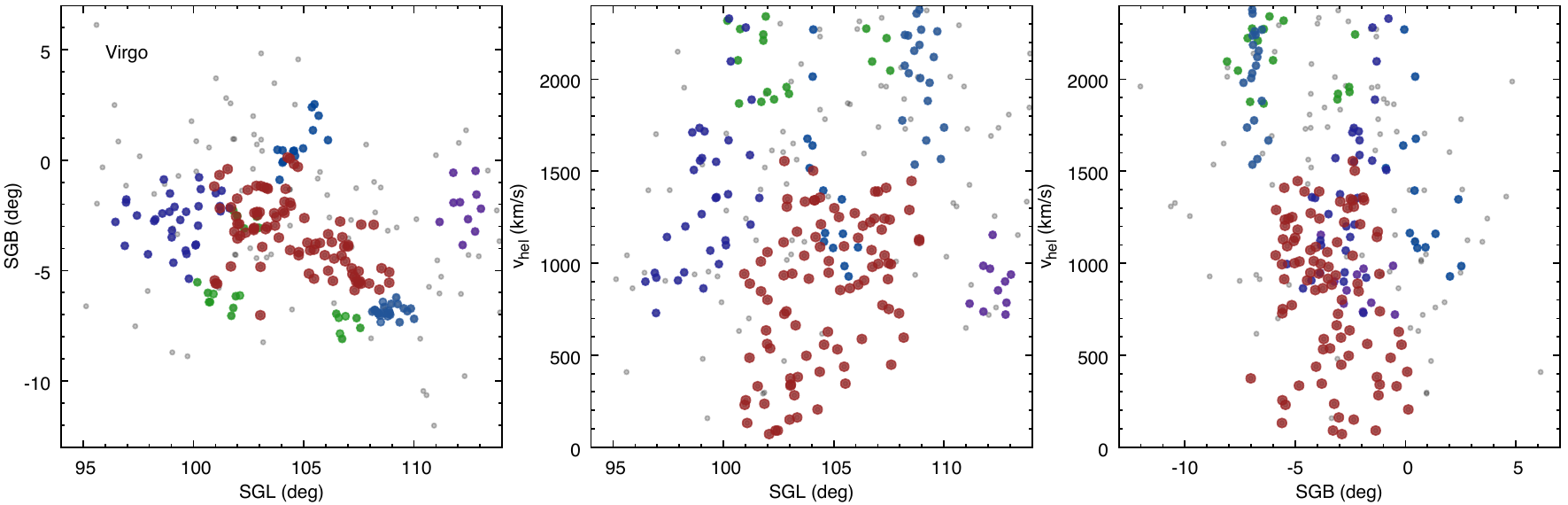}
	\caption{Virgo cluster region in 2MRS data set. Panels are the same as in Fig.~\ref{fig:coma}. Red points show the central component of the Virgo cluster. Bluish points show the detected systems with ten or more members. Green points show detected groups with four to seven member galaxies. Grey points show all other galaxies in Virgo cluster region.}
	\label{fig:virgo}
\end{figure*}

There are several well known clusters (the existence of the clusters is known but not the properties of the clusters) in the local Universe. In Figs.~\ref{fig:coma} and \ref{fig:virgo} we show the Coma and Virgo cluster regions, respectively. In case of Coma cluster (see Fig.~\ref{fig:coma}) the central region (red points) is very well detected by our Bayesian group finder containing 84 galaxies. Around the Coma cluster, there is one relatively large system with 15 galaxies (blue points on the Fig.~\ref{fig:coma}) and several smaller systems with three to four galaxies marked as green points. Based on Fig.~\ref{fig:coma}, the Coma cluster is well detected by our Bayesian group finder and the fragmentation of the Coma cluster outer regions into several systems is justified. The individual smaller systems are clearly separated in sky plane and/or in velocity space. Using the Eq.~(\ref{eq:mass}), the estimated mass of the central component of Coma cluster is $M_\mathrm{vir} = 1.0\times 10^{15}~M_\odot$, which is the mass inside the virial radius $R_{200} = 2.1~\mathrm{Mpc}$. This mass is somewhat lower than estimated mass $M_\mathrm{vir} = 1.8\times 10^{15}~M_\odot$ by \citet{2015AJ....149...54T}. However, the mass is very sensitive to the measured velocity dispersion of the system. In case of \citet{2015AJ....149...54T} the velocity dispersion is measured using all galaxies in a wider window in velocity space, which yields higher velocity dispersion. Hence the mass difference is purely described by the difference in velocity dispersion. Using dynamical mass estimates, the mass of the Coma cluster is estimated to be $M_\mathrm{vir} = 0.93\pm 0.04\times 10^{15}~M_\odot$ \citep{2003AJ....126.2152R} and $M_\mathrm{vir} = 1.24\pm 0.46\times 10^{15}~M_\odot$ \citep{2003MNRAS.343..401L}. Using weak lensing analysis the estimated Coma cluster mass is $M_\mathrm{vir} = 1.2_{-0.3}^{+1.3}\times 10^{15}~M_\odot$ \citep{2014ApJ...784...90O}. These values are much closer to our estimate, considering that our mass estimate includes only the central component of the Coma cluster, hence the total mass of Coma cluster is even higher.

	\begin{figure}
		\includegraphics[width=88mm]{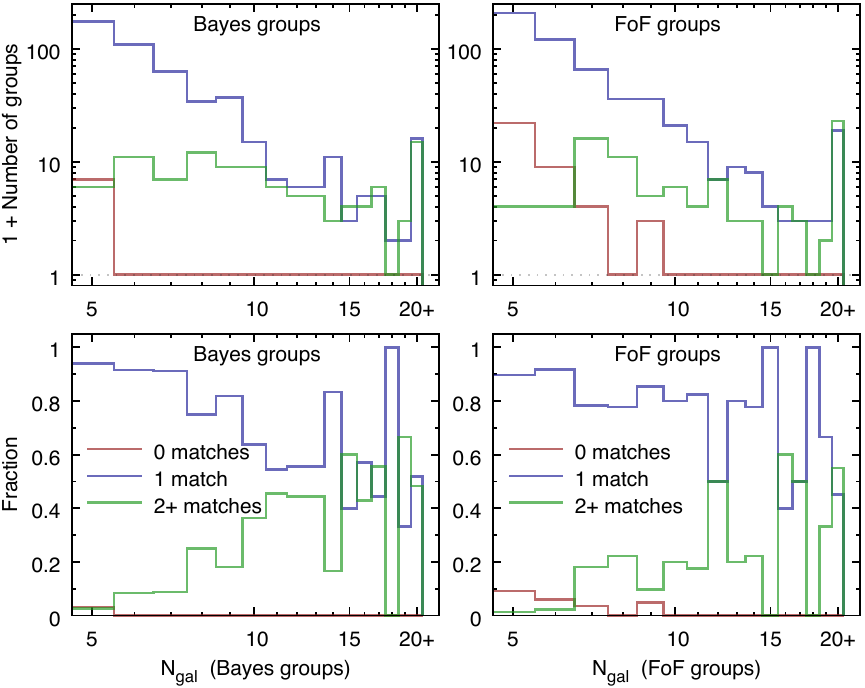}
		\caption{Upper panels: the distribution of groups richnesses ($N_\mathrm{gal}$) with zero (red line), one (blue line), and two or more (green line) matches in comparison group catalogue. Lower panels: the fraction of groups with zero (red line), one (blue line), and two or more (green line) matches as a function of group richness. Left panels show the distribution and fraction for Bayes groups with matches in the FoF group catalogue. Right panels show the distribution and fraction for FoF groups with matches in the Bayes group catalogue.}
		\label{fig:n_matches} 
	\end{figure}

\object{Virgo cluster} is one of our closest nearby massive galaxy clusters, which makes it very extended system on the sky plane. Because of its proximity, Virgo cluster has received a lot of attention (see e.g. \citealt{2012ApJS..200....4F} and \citealt{2015AJ....149...54T} and discussions therein) and it is well known that the cluster is composed of multiple components \citep[see e.g.][]{1994Natur.368..828B}. Figure~\ref{fig:virgo} shows the Virgo cluster region in 2MRS data set together with the detected systems using our Bayesian group finder. The central system in Virgo cluster is well detected and it includes 90 galaxies. There are four other detected groups close to the Virgo central component that contain 10 to 32 galaxies (bluish points in Fig.~\ref{fig:virgo}). The complicated structure of the Virgo cluster is well discussed in \citet{2015AJ....149...54T}. Despite the complicated nature of the Virgo cluster, our Bayesian group finder detects the main components of the cluster reasonably well. The estimated summed mass of galaxy systems detected in Virgo cluster region is $M_\mathrm{vir} = 4.8\times 10^{14}~M_\odot$, which is slightly lower than $M_\mathrm{vir} = 7$--$8\times 10^{14}~M_\odot$ estimated using the virial and infall mass estimates \citep{2005ApJ...635L.113M, 2015AJ....149...54T} or $M = 6.3\pm 0.8\times 10^{14}~M_\odot$ using the first turn around radius \citep{2017ApJ...850..207S}. Once again, the summed mass of our Bayesian groups does not reflect the total mass of the Virgo cluster and our summed mass is expected to be slightly lower than the total mass of the Virgo cluster.

Using two examples, Coma and Virgo clusters, we demonstrated that our Bayesian group finder works well for these systems and it detects the main components of the two known large clusters. Visual assessment of all other rich systems yields to the same conclusion that the detected systems are reasonable. However, the detection of the systems is only one aspect of the problem. The second aspect is how well the detected systems parameters can be recovered. Our current Bayesian group finder is not specifically tuned to recover the true parameters of the systems, nevertheless the mass of the Virgo and Coma cluster is in the same order of magnitude as reported in the literature. It requires a special study and the use of mock data sets in order to analyse how reliably our Bayesian methodology recovers the true systems and how to improve the group parameter estimation in our Bayesian group finder.

\subsection{Comparison of Bayesian and FoF groups}
\label{sec:comparison}

	In this section we compare the constructed Bayes group catalogue with the FoF group catalogue published by \citet{2016A&A...588A..14T}. Both group catalogues are based on exactly the same data, hence we can make a direct comparison between these two catalogues. We limit ourselves with groups closer than 300~Mpc in both catalogues, which gives 7755 groups with at least two members in our Bayes group catalogue and 6251 groups in the FoF group catalogue. Excluding galaxy pairs, it gives 1933 and 2525 groups in Bayes and FoF group catalogues, respectively. Although there are more groups in the Bayes group catalogue, majority of them are galaxy pairs. Considering only groups with more than five members, then the numbers are more similar, 391 and 431 groups in Bayes and FoF catalogues, respectively. Hence, there are fewer richer groups in our Bayes group catalogue, but the difference is less than 10 per cent. As can be seen from Fig.~\ref{fig:ngal_vs_dist} the distribution of groups as a function of distance in both catalogues is similar, which is expected since the grouping parameter (linking length in the FoF or object size $r_0$ in the current Bayesian group finder) as a function of distance is chosen based on the same considerations (see Section~\ref{sec:2mrs_settings}).

	Comparing the number of galaxies in groups, then there are 50 and 46 per cent of galaxies in Bayes and FoF groups, respectively. In overall, our Bayesian group finder leaves 4 per cent less unclustered galaxies than the FoF group finder. Most of the difference comes from galaxy pairs, there are more galaxy pairs in the Bayes group catalogue, which is related with the choice of the data term in our model. Looking at the separation of galaxy pairs on the sky plane and in redshift space, then the separation of galaxy pairs on the sky plane is on average 28 per cent larger for Bayes groups compared with FoF groups. The separation of galaxy pairs in redshift space is roughly the same in both catalogues. This can be explained by the object shape properties in our Bayesian group finder, the maximum object size on the sky plane is 1.5 times larger than the linking length in the FoF group finder.

	To compare groups in both catalogues one-to-one basis, we first compare how groups in one catalogue are detected in another catalogue. For this we take all members of one group and look between how many groups these member galaxies are distributed in another catalogue, that is for each group we get the number of matching groups in another catalogue. In this analysis we only use groups with at least three galaxies. The results are shown in Fig.~\ref{fig:n_matches}, where left panels show the results for Bayes groups and right panels for FoF groups. In both catalogues there is a small number of groups that are undetected using another method (red lines on the figure). All these groups have less than ten galaxies and the fraction of these groups is less than 10 per cent for each group richness bin. Hence, all groups with at least ten galaxies have a detection in both catalogues and majority of groups with less than ten galaxies have a counterpart in another catalogue. Green lines in Fig.~\ref{fig:n_matches} show the distribution and fraction of groups with two or more matches in comparison group catalogue. The fraction of these groups increases with group richness reaching to about 0.5. Hence, majority of smaller groups and about half of richer groups have only one counterpart in the comparison catalogue.
	
	\begin{figure}
		\includegraphics[width=88mm]{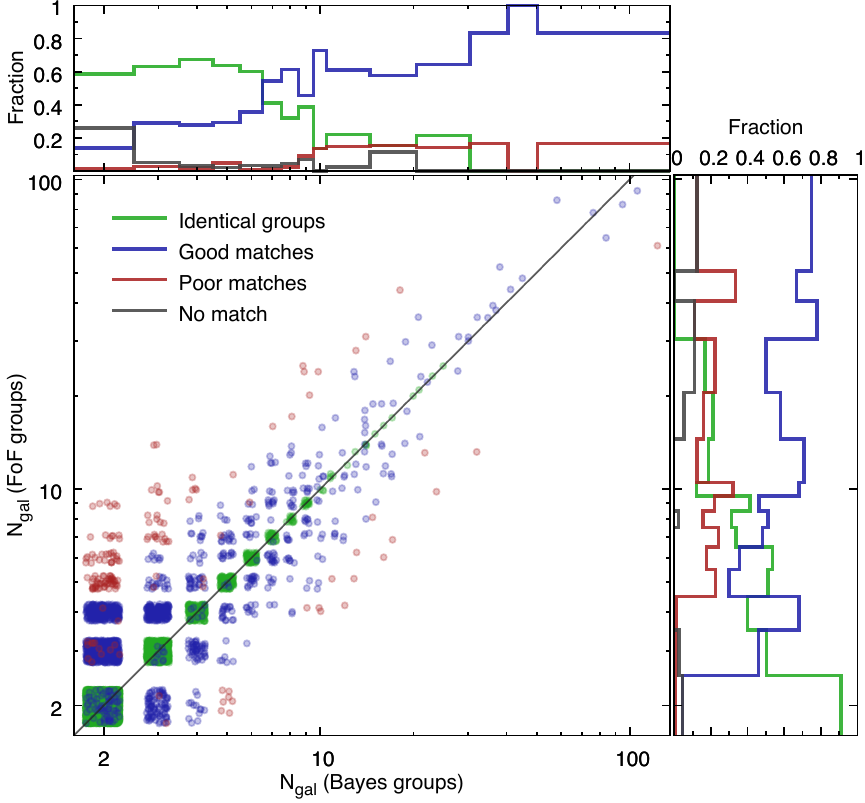}
		\caption{The comparison of group richness ($N_\mathrm{gal}$) in Bayes and FoF groups. Green points indicate identical groups, blue ones show groups where at least half of the galaxies are common in FoF and Bayes groups, and red ones are groups that differ significantly in Bayes and FoF group catalogues. For illustrative purposes, each point have a small random scatter around the true value. Black line shows the one-to-one relationship. Although some of the groups contain the same number of galaxies, they are not identical. The right and upper side panels show the fractions as a function of group richness.}
		\label{fig:ngal_ngal}
	\end{figure}

	Another question we may ask is how does the group richness depend on the grouping method (Bayesian or FoF). For this we have to match the groups in both catalogues. To find a match for one Bayes group in the FoF group catalogue (and vice versa), we require that the matching group have the highest number of common galaxies in matched group catalogue and vice versa. This ensures that the found matches are independent of which catalogue was taken as the base catalogue for matching. Around 80 per cent of all Bayes groups have a well defined match in the comparison FoF group catalogue. If we only consider groups with at least three galaxies in Bayes groups, then 96 per cent of all Bayes groups have a clear match in the FoF catalogue. Hence, only a small fraction of Bayes groups with at least three galaxies does not have a clear match in the comparison FoF catalogue. This shows that the same groups are detected regardless of the used (Bayesian of FoF) grouping method. In Fig.~\ref{fig:ngal_ngal} we show the richness of matched groups in Bayes and FoF group catalogues. Among those groups that have clear match in both catalogues, around three quarters (74 per cent) of groups are identical (contain exactly the same galaxies) in Bayes and FoF group catalogues and for 97 per cent of groups at least half of the galaxies are common in both catalogues. Figure~\ref{fig:ngal_ngal} also shows that there is small tendency that Bayes groups contain on average less galaxies than FoF groups (there are slightly more points above the one-to-one relationship in Fig.~\ref{fig:ngal_ngal}), but this concerns mostly smaller groups. This is partially explained by the fact that in the FoF group catalogue the group membership refinement (see Section~\ref{sec:group_refinement}) was only applied for groups with at least five galaxies.

\begin{figure}
			\includegraphics[width=88mm]{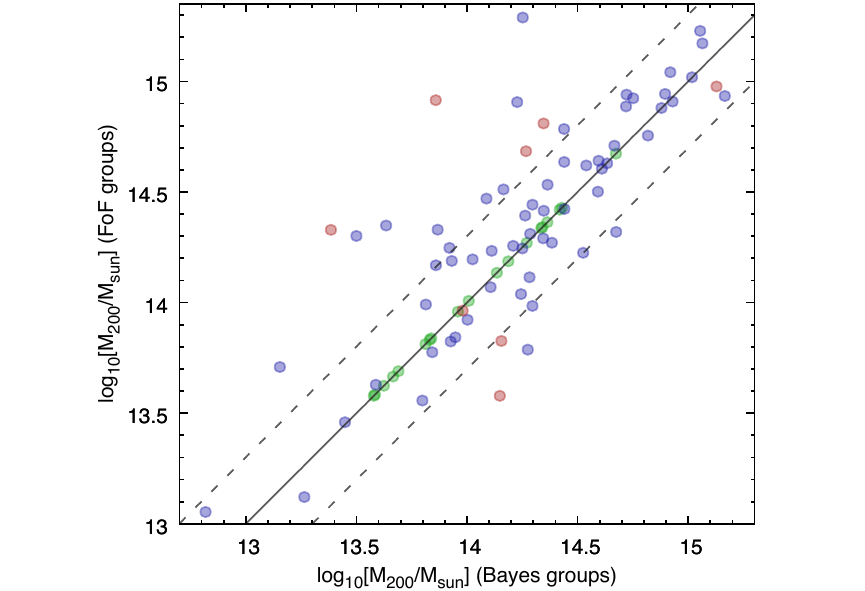}
			\caption{Group masses in Bayes and FoF group catalogues for groups with at least ten galaxies in both catalogues. Green points indicate identical groups, blue ones show groups where at least half of the galaxies are common in FoF and Bayes groups, and red ones are groups that differ significantly. Black line shows the one-to-one relationship and dashed lines indicate the region where mass difference is less than two times.}
			\label{fig:mass_mass}
		\end{figure}

	In Fig.~\ref{fig:mass_mass} we compare the masses of groups in Bayes and FoF group catalogues. Since mass estimation is rather uncertain for poor groups (with small number of galaxies), we only use groups that have at least ten galaxies in both catalogues. The mass estimation algorithm is the same in both catalogues, the only difference is the group membership. Figure~\ref{fig:mass_mass} shows a scatter in estimated group masses around the one-to-one relationship. Group masses in Bayes and FoF catalogues for most of the cases differ less than two times. There is small tendency that masses in the FoF group catalogue are slightly higher than masses in the Bayes group catalogue, which comes from the sensitivity of the mass estimator to the group boundaries (group membership). As groups in the Bayes group catalogue contain (statistically) less galaxies (see Fig.~\ref{fig:ngal_ngal}) they are also slightly smaller (group membership differs mostly on the outer edges of groups), hence their estimated mass is slightly smaller.

		In general, the groups extracted using Bayesian (this work) or FoF group finder \citep{2016A&A...588A..14T} are very similar and majority of the groups are identical. On the one hand this means that most of the groups in the local Universe are well defined and easily detectable. On the other hand we can infer from it that with reasonable grouping parameters the two seemingly different methods provide very similar results. Considering that the FoF grouping method is well tested using simulated mock data \citep{2014MNRAS.441.1513O,2015MNRAS.449.1897O,2018MNRAS.475..853O}, the Bayesian grouping methodology in its simplest form performs very well and it opens up the possibility to study the potential of the proposed Bayesian methodology even further. The differences between Bayes and FoF groups are in details, which will be briefly analysed in the next section.

		\begin{figure}
			\includegraphics[width=88mm]{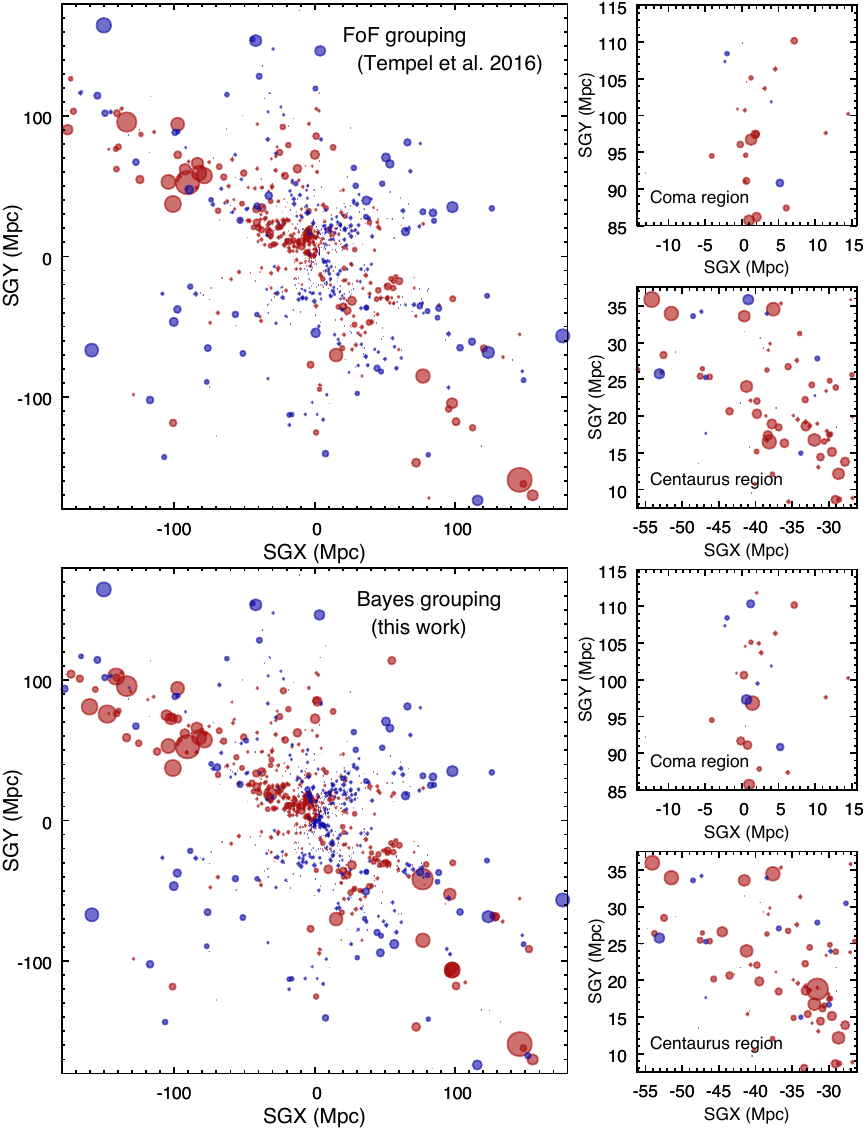}
			\caption{Radial peculiar velocities of galaxies in a supergalactic slice with 15~Mpc thickness. Radial peculiar velocities are grouped (averaged) using the groups obtained with a FoF algorithm (upper panels) and the Bayesian group finder (lower panels). A red dot means that the peculiar velocity is positive while a blue dot means that it is negative, the dot size is proportional to the absolute value of the radial peculiar velocity. Smaller panels on the right-hand side show two specific regions, Coma and Centaurus cluster regions, to highlight some of the differences in detail.}
			\label{fig:wf_constraints}
		\end{figure}
	
		\begin{figure*}
			\centering
			\includegraphics[width=91mm]{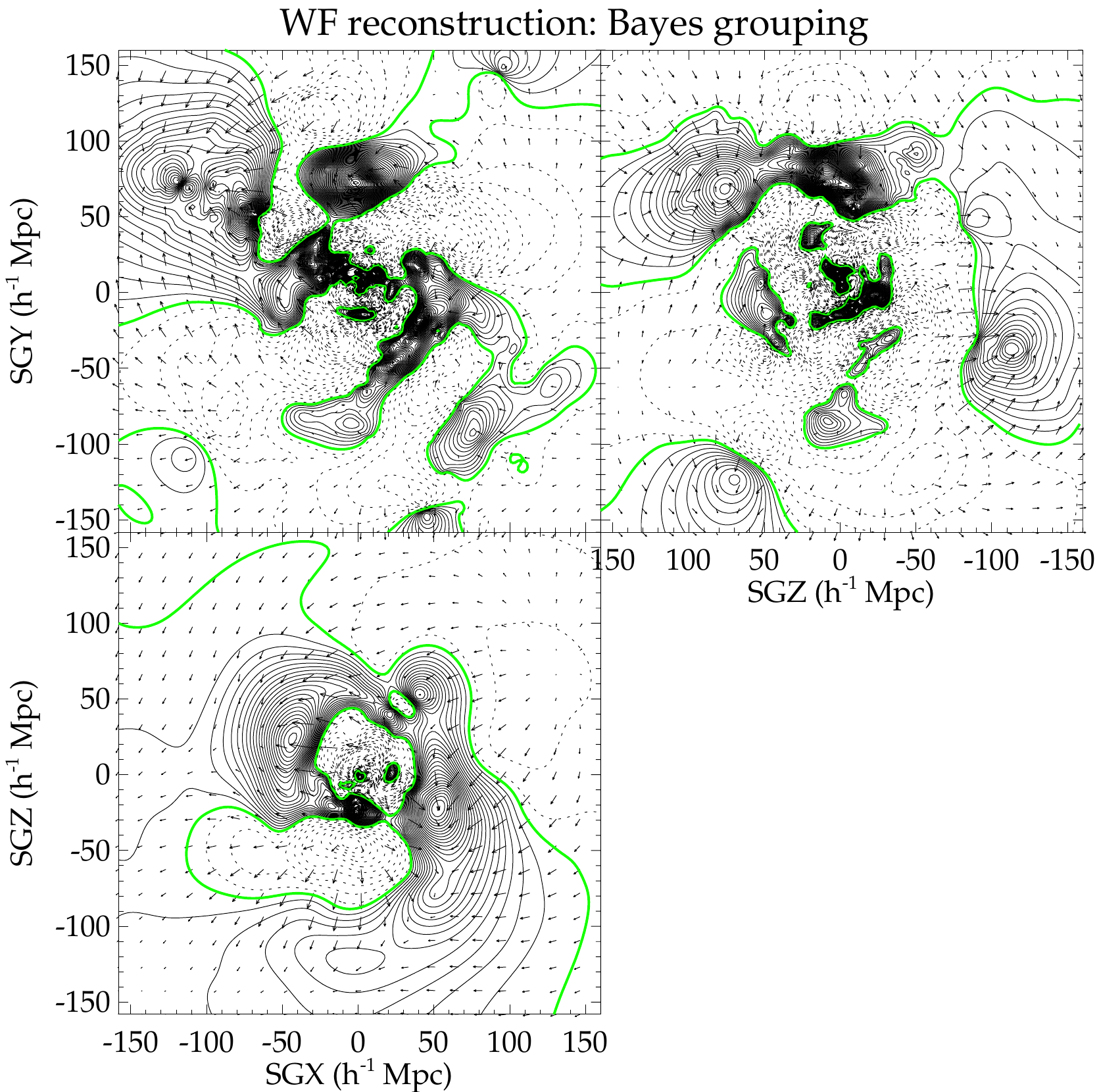}
			\includegraphics[width=91mm]{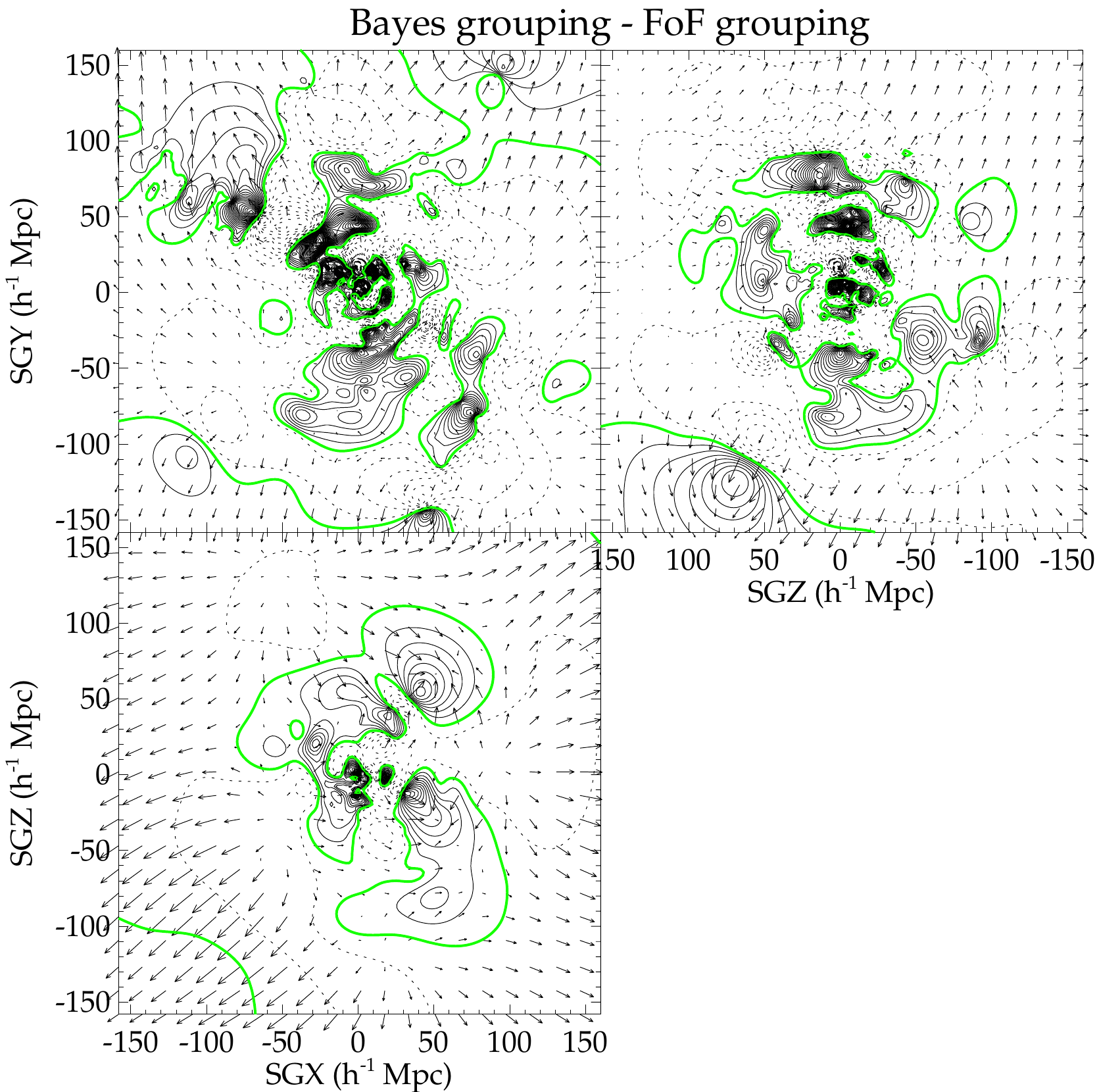}
			\caption{Left-hand panels: supergalactic slices of the reconstructed velocity (arrows) and over-density (contours) fields of the local Universe. The green colour stands for the mean density field, dashed contours are under-density regions while solid contours are over-density regions. Right-hand panels: the residual between two reconstructions obtained with Bayesian and FoF grouping schemes. See also fig~5 in \citet{2017MNRAS.469.2859S}.}
			\label{fig:wf_fields}
		\end{figure*}

\subsection{Application of Bayesian groups for constrained simulations}

	One application, where group catalogues of local Universe are utilised, is the construction of constrained simulations. In \citet{2017MNRAS.469.2859S} Wiener Filter technique \citep{1995ApJ...449..446Z, 1999ApJ...520..413Z} is applied to the catalogue of radial peculiar velocities derived from the cosmicflows project \citep{2013AJ....146...86T} in order to reconstruct the density and velocity field in the local Universe. To suppress the non-linear motions, the radial peculiar velocities are grouped using the FoF galaxy groups published by \citet{2016A&A...588A..14T}. As follows, we will do a brief comparison how the reconstruction of the local Universe changes if we use the Bayes group catalogue instead of the FoF group catalogue.

	In Fig.~\ref{fig:wf_constraints} we show the peculiar velocity constraints for the FoF (upper panel) and Bayes (lower panel) groups. Radial peculiar velocities are grouped (an average value is taken) according to the used group catalogue. Detailed descriptions are given in \citet{2017MNRAS.469.2859S}. Overall, the two grouping schemes exhibit catalogues in agreement with each other, the constraints are quite similar. However, zooming on a particularly dense region, like the Coma cluster area or the Centaurus cluster region, differences are more pronounced. To see how these differences affect the reconstruction, we show in Fig.~\ref{fig:wf_fields} the over-density and velocity field using the Bayesian grouping (left panels) and the residual between reconstructions (right panels) obtained with Bayesian and FoF grouping. Although the Bayes and FoF group catalogues are rather similar, there are clear differences between the reconstructed density and velocity fields (see right panels in Fig.~\ref{fig:wf_fields}). These differences are of the same order as differences between FoF grouping and more aggressive grouping algorithm (see details and fig~5 in \citealt{2017MNRAS.469.2859S}). Although the differences in mean density and velocity fields are small, they change the mass of the large clusters by about at least 5 per cent on average. When analyzing the specific clusters in the local Universe, \citet{2018MNRAS.476.4362S} conclude that the grouping scheme used to remove non-linear motions in the catalogues that constrain the simulations affect the quality of the numerical clusters. These comparisons emphasise that the choice of grouping method is important in order to suppress the non-linear motions. Hence, it is important to improve the available grouping techniques for future studies.

\section{Conclusions}
\label{sec:conclusions}

   In this study we proposed a new cluster pattern detection algorithm for the identification of galaxy groups and clusters in cosmological galaxy redshift surveys. The proposed method is based on object point processes and follows the methodology described in \cite{Stoica:07}. In the method we use the marked point process in order to model the cluster pattern visible in galaxy redshift surveys. The observed galaxy data is only used to assign probabilities for marked points in our point process model. The marked point process is modelled in a Bayesian framework, which allows us to study the detected cluster pattern in a probabilisitic way. As a result of this, the detection of galaxy groups and clusters in the Bayesian group finder is probabilistic, which can be taken into account in practical applications. The proposed approach is complementary to the currently available methods for galaxy group and cluster detection, where the galaxy data is used directly (no modelling of underlying cluster pattern).
   
   We applied the proposed Bayesian group finder to the 2MRS data set and extracted galaxy groups from the data. We compared our group catalogue with galaxy group catalogue constructed using the FoF algorithm \citep{2016A&A...588A..14T} and showed that the two group catalogues are very similar and they differ only in details. Majority of the groups (70 per cent) are identical in both catalogues and more than 90 per cent of them have a clear counterpart in another catalogue. Although the differences between two group catalogues are in details, they might have significant effect on the specific applications. Following the approach described in \citet{2017MNRAS.469.2859S}, we used the constructed Bayes group catalogue to reconstruct the density and velocity fields in the local Universe. This comparison emphasises that the grouping of galaxies (in order to suppress non-linear motions) have significant effect on the reconstructed galaxy clusters. This emphasises the need for a reliable grouping algorithm and encourages to continue the work on the proposed Bayesian group finder in order to improve it.
   
   Our proposed group finder has two distinct components. The basis of the Bayesian group finder is an object point process with interactions that we use to generate the probabilistic group detection field (visit map for group detection). This step is fully determined by the marked point process theory and the detection of cluster pattern can be mainly influenced by the assigned probabilities that depend on the locations and parameters of observed galaxies. The second component of the proposed group finder involves the post-processing of the probabilistic group detection field. Here, various approaches can be taken and depending on the specific study different methods can be the preferred ones. In this paper we proposed an approach that is parameter free and simple to implement. In general, the detected groups of galaxies are in the peaks of the obtained visit map, which is the outcome of a number of simulations of the detection model. It requires a dedicated study in order to analyse various post-processing possibilities for galaxy group and cluster extraction from the probabilistic visit map data.
   
   In this paper we introduced the concept of Bayesian group finder based on marked point processes. The application to the 2MRS data set and comparisons with the published FoF catalogue show that the new methodology for group detection is feasible. Application of the proposed method to a simulated mock data set shows that the method in its simplest form perform equally well in terms of contamination and incompleteness with other available methods for group and cluster detection (see Appendix~\ref{app:mock} and \citealt{2018arXiv180603199W}). The Bayesian approach introduced in this paper provides ``for free'' additional information such as the probabilities that a point or two points (e.g. galaxies) in the observation domain belong to the cluster pattern. These supplementary tools allow the construction of tests and techniques to validate and to refine the detection result. In a forthcoming study we will apply the proposed methodology to the simulated mock data to test the full potential of the proposed methodology.

	In current study we proposed the object point process methodology for the group detection in galaxy redshift surveys. However, our model can be adapted to any data sets, where group or clump detection is required. In general, our proposed model is meant to detect cluster pattern that does not directly depend on the used data set. The observational data in our model is only used to define the data energy term (see Section~\ref{sec:dataterm}), which can be easily adjusted according to the used data set. Additional advantage of our approach is that it is straightforward to include observational uncertainties in our model, which only requires the modification of the data energy term in the model. It is especially promising for the photometric redshift surveys such as J-PAS \citep{2014arXiv1403.5237B}, where the full photometric redshift posterior can be taken into account in our Bayesian group finder.

\begin{acknowledgements}
      We thank the Referee for her/his suggestions how to improve the manuscript. We thank Enn Saar and Stefan Gottl\"ober for valuable comments. Part of this work was supported by institutional research funding \mbox{IUT26-}2 and \mbox{IUT40-2} of the Estonian Ministry of Education and Research. We acknowledge the support by the Centre of Excellence ``Dark side of the Universe'' (TK133) and by the grant MOBTP86 financed by the European Union through the European Regional Development Fund. JS acknowledges support from the Centre National d'\'etudes spatiales (CNES) research fellowship and from the l'Or\'eal-UNESCO ``pour les femmes et la science'' fellowship programs. Part of the work of RS was supported by a grant of the Romanian Ministry of National Education and Scientific Research, RDI Programme for Space Technology and Advanced Research -- STAR, project number 513.\\
	  The CosmoSim database used in this paper is a service by the Leibniz-Institute for Astrophysics Potsdam (AIP). The MultiDark database was developed in cooperation with the Spanish MultiDark Consolider Project CSD2009-00064. The authors gratefully acknowledge the Gauss Centre for Supercomputing e.V. (www.gauss-centre.eu) and the Partnership for Advanced Supercomputing in Europe (PRACE, www.prace-ri.eu) for funding the MultiDark simulation project by providing computing time on the GCS Supercomputer SuperMUC at Leibniz Supercomputing Centre (LRZ, www.lrz.de).
\end{acknowledgements}


\appendix

\section{Description of the catalogue}
\label{app:cat}

The catalogue of galaxy groups and clusters consists of two tables. The first table lists galaxies that were used to generate the group catalogues, the second describes the group properties. The catalogues are available at \url{http://cosmodb.to.ee} and the catalogues will also be made available through the Strasbourg Astronomical Data Centre (CDS).

\subsection{Galaxy table}

The galaxy table contains the following information (column numbers are given in square brackets):
\begin{enumerate}
 \item{[1]\,\texttt{pgcid} --} identification number in PGC (principal galaxy catalogue);
 \item{[2]\,\texttt{groupid\_bgf} --} primary group or cluster id (given in the present paper) the galaxy belongs to; may be different from \texttt{icl1} due to the group membership refinement procedure;
 \item{[3]\,\texttt{ngal\_bgf} --} richness (number of members) of the primary group or cluster the galaxy belongs to;
 \item{[4]\,\texttt{groupdist} --} comoving distance to the centre of the primary group or cluster the galaxy belongs to, in units of Mpc, calculated as an average over all galaxies within the group or cluster;
 \item{[5]\,\texttt{zobs} --} observed redshift (without the CMB correction);
 \item{[6]\,\texttt{zcmb} --} redshift, corrected to the CMB rest frame;
 \item{[7]\,\texttt{zerr} --} error of the observed redshift;
 \item{[8]\,\texttt{dist} --} comoving distance in units of Mpc;
 \item{[9--10]\,\texttt{raj2000, dej2000} --} right ascension and declination (deg);
 \item{[11--12]\,\texttt{glon, glat} --} Galactic longitude and latitude (deg);
 \item{[13--14]\,\texttt{sglon, sglat} --} supergalactic longitude and latitude (deg);
 \item{[15--17]\,\texttt{xyz\_sg} --} supergalactic cartesian coordinates in units of Mpc;
 \item{[18]\,\texttt{mag\_ks} --} Galactic-extinction-corrected $K_s$ magnitude as given in the source catalogue;
 \item{[19]\,\texttt{groupid\_fof} --} group or cluster id in the FoF group catalogue \citep{2016A&A...588A..14T};
 \item{[20]\,\texttt{ngal\_fof} --} richness (number of members) of the group or cluster the galaxy belongs to, based on the FoF group catalogue \citep{2016A&A...588A..14T};
 \item{[21]\,\texttt{n\_cl} --} number of groups and clusters the galaxy is associated with;
 \item{[22—26]\,\texttt{icl1..5} --} identification numbers of group or cluster the galaxy is associated with;
 \item{[27]\,\texttt{p\_field} --} probability that the galaxy is a field galaxy;
 \item{[28—32]\,\texttt{p1..5} --} probabilities that the galaxy belongs to the specified group or cluster.
\end{enumerate}

\subsection{Group table}

The group and cluster table contains the following information (column numbers are given in square brackets):
\begin{enumerate}
 \item{[1]\,\texttt{groupid} --} group or cluster identification number, given in the present paper;
 \item{[2]\,\texttt{ngal} --} richness (number of members) of the group;
 \item{[3--4]\,\texttt{raj2000, dej2000} --} right ascension and declination of the group centre (deg);
 \item{[5--6]\,\texttt{glon, glat} --} Galactic longitude and latitude of the group centre (deg);
 \item{[7--8]\,\texttt{sglon, sglat} --} supergalactic longitude and latitude of the group centre (deg);
 \item{[9]\,\texttt{zcmb} --} CMB-corrected redshift of the group, calculated as an average over all group or cluster members;
 \item{[10]\,\texttt{groupdist} --} comoving distance to the group centre (Mpc);
 \item{[11]\,\texttt{sigma\_v} --} rms deviation of the radial velocities ($\sigma_v$ in physical coordinates, in \mbox{km\,s$^{-1}$});
 \item{[12]\,\texttt{sigma\_sky} --} rms deviation of the projected distances in the sky from the group centre ($\sigma_\mathrm{sky}$ in physical coordinates, in Mpc), $\sigma_\mathrm{sky}$ defines the extent of the group in the sky;
 \item{[13]\,\texttt{r\_max} --} distance (in Mpc) from group centre to the farthest group member in the plane of the sky;
 \item{[14]\,\texttt{mass\_200} --} estimated mass of the group assuming the NFW density profile (in units of $10^{12}M_\odot$);
 \item{[15]\,\texttt{r\_200} --} radius (in kpc) of the sphere in which the mean density of the group is 200 times higher than the average of the Universe;
 \item{[16--18]\,\texttt{xyz\_sg} --} supergalactic cartesian coordinates in units of Mpc.
\end{enumerate} 

\begin{figure*}
	\sidecaption
	\includegraphics[width=120mm]{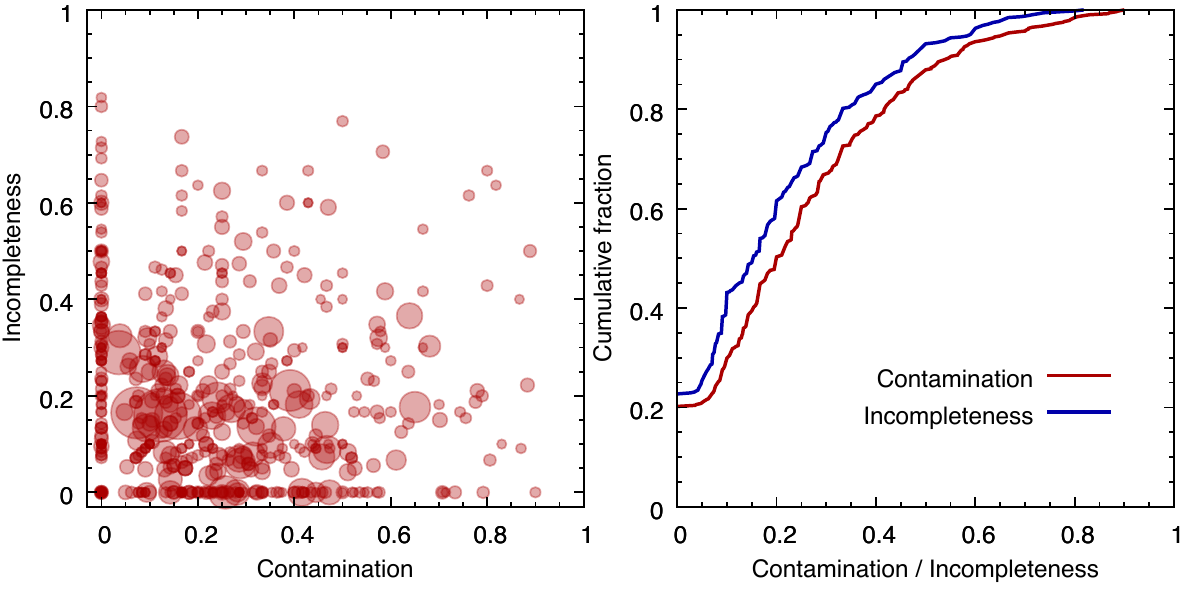}
	\caption{Left-hand panel shows the contamination versus incompleteness for groups with at least ten members. Point size is proportional to the number of true members in a group. Majority of the large groups and clusters are located in bottom left corner with low contamination and incompleteness. This figure should be compared with results presented in \citet{2018arXiv180603199W}. Right-hand panel shows the cumulative distribution of contamination and incompleteness for groups shown in the left-hand panel. For about 20 per cent of the groups the contamination (red line) or incompleteness (blue line) is exactly zero.}
	\label{fig:mock}
\end{figure*}

\section{Application to simulated mock data}
\label{app:mock}

We constructed a simulated mock catalogue based on the MultiDark simulation\footnote{\url{https://www.cosmosim.org}} in a 1~$h^{-1}$Gpc box \citep{2013AN....334..691R, 2016MNRAS.457.4340K}. For mock catalogue we used SAG semi-analytic galaxies attached to the MultiDark MDPL2 simulation as described in \citet{2018MNRAS.474.5206K}. To have a reasonable sample size, the mock catalogue was constructed using a $(250~h^{-1}\mathrm{Mpc})^3$ sub-region from the full box. We selected all semi-analytic galaxies brighter than $r$-band absolute magnitude $-19.5$~mag. Observer was located in one corner of the box and peculiar velocities of galaxies along the line of sight were added to galaxies to mimic the Fingers-of-God effect.

The true membership of groups and clusters were determined based on the dark matter halo catalogue published together with MultiDark simulation. The largest cluster in our mock catalogue contains 94 semi-analytic galaxies.

We applied the Bayesian group finder to simulated mock data using the same model parameters as we used for the 2MRS data set. The only exception was the base object radius, which was determined based on the mean separation of galaxies in the mock catalogue. Since our mock catalogue is a volume-limited we used distance independent object radius in our model. The object radius for mock data set had a uniform law over $[0.2,0.5]~h^{-1}$Mpc.

To analyse the contamination and incompleteness of the constructed groups we followed \citet{2018arXiv180603199W}, where the contamination ($C$) and incompleteness ($I$) are defined as
\begin{eqnarray*}
	C &=& \frac{N_{\mathrm{Bayes,non-mem}}}{N_\mathrm{Bayes}},\\
	I &=& \frac{N_{\mathrm{non-Bayes,mem}}}{N_\mathrm{mem}},
\end{eqnarray*}
where $N_\mathrm{mem}$ is the number of true members, $N_\mathrm{Bayes}$ is the number of members in extracted Bayes groups, $N_{\mathrm{Bayes,non-mem}}$ is the number of interlopers (members in Bayes groups but not true members) and $N_{\mathrm{non-Bayes,mem}}$ is the number of missing true members in Bayes groups. Contamination and incompleteness are defined between 0 and 1. For a perfectly extracted groups with no interlopers and all true members included, $C=0$ and $I=0$.

For the analysis that follows we used only groups with at least ten true members. For each dark matter halo the matching Bayes group is the one with highest number of common members. Hence, for each true group or cluster we have only one matching group in Bayes group catalogue. The match exists for each true group.

Figure~\ref{fig:mock} shows the contamination versus incompleteness for extracted Bayes groups. Most of the groups are located in bottom left corner with low contamination and incompleteness. The contamination and incompleteness is roughly the same as in other methods analysed in \citet{2018arXiv180603199W}. Hence, our Bayesian group finder perform equally well with other available methods. Analysis presented in \citet{2018arXiv180603199W} suggests that most of the contamination and incompleteness comes from the redshift-space distortions, not from the used grouping method.

Right-hand panel in Fig.~\ref{fig:mock} shows the cumulative distribution of contamination and incompleteness. For half of the groups both, the contamination and incompleteness, are less than 20 per cent. However, for 10 per cent of detected groups the contamination or incompleteness is higher than 0.5. As can be seen from left-hand panel of Fig.~\ref{fig:mock} most of high contamination and incompleteness groups are poor systems. Large groups and clusters are relatively well detected in the Bayesian group finder.

\end{document}